\documentclass[letterpaper]{article} 

\usepackage[]{aaai24}  
\usepackage{times}  
\usepackage{helvet}  
\usepackage{courier}  
\usepackage[hyphens]{url}  
\usepackage{graphicx} 
\usepackage{natbib}  
\usepackage{caption} 
\usepackage{algorithm}
\usepackage{algorithmic}
\usepackage{subcaption,graphicx}

\usepackage{newfloat}
\usepackage{listings}
\DeclareCaptionStyle{ruled}{labelfont=normalfont,labelsep=colon,strut=off} 
\floatstyle{ruled}
\newfloat{listing}{tb}{lst}{}
\floatname{listing}{Listing}
\title{Do Generative AI Models Output Harm while Representing Non-Western Cultures: Evidence from A Community-Centered Approach}

\author{
    Sourojit Ghosh\textsuperscript{\rm 1}\equalcontrib, Pranav Narayanan Venkit\textsuperscript{\rm 2}\equalcontrib, Sanjana Gautam\textsuperscript{\rm 2}\equalcontrib,\\
    Shomir Wilson\textsuperscript{\rm 2}, Aylin Caliskan\textsuperscript{\rm 1}
}
\affiliations{
    \textsuperscript{\rm 1}University of Washington, Seattle, Washington, USA\\
    \textsuperscript{\rm 2}Pennsylvania State University, University Park, Pennsylvania, USA\\
    \{ghosh100@uw, pranav.venkit@psu, sanjana.gautam@psu, shomir@psu, aylin@uw\}.edu

}

\begin{document}

\maketitle

\begin{abstract}
Our research investigates the impact of Generative Artificial Intelligence (GAI) models, specifically text-to-image generators (T2Is), on the representation of non-Western cultures, with a focus on Indian contexts. Despite the transformative potential of T2Is in content creation, concerns have arisen regarding biases that may lead to misrepresentations and marginalizations. Through a community-centered approach and grounded theory analysis of 5 focus groups from diverse Indian subcultures, we explore how T2I outputs to English prompts depict Indian culture and its subcultures, uncovering novel representational harms such as exoticism and cultural misappropriation. These findings highlight the urgent need for inclusive and culturally sensitive T2I systems. We propose design guidelines informed by a sociotechnical perspective, aiming to address these issues and contribute to the development of more equitable and representative GAI technologies globally. Our work also underscores the necessity of adopting a community-centered approach to comprehend the sociotechnical dynamics of these models, complementing existing work in this space while identifying and addressing the potential negative repercussions and harms that may arise when these models are deployed on a global scale.
\\
\\
\textbf{Disclaimer: This paper contains images that may be considered derogatory or offensive; reader discretion is advised.}
\\
\\
\textit{This is the pre-peer reviewed version, which has been accepted at the 7th AAAI/ACM Conference on AI, Ethics, and Society, Oct. 21–23, 2024, California, USA.}
\end{abstract}

\section{Introduction}
Innovations into Generative Artificial Intelligence (GAI) tools such as text-to-image generators (T2Is) have revolutionized the way we create content, such as producing realistic images/videos. This progress has seen a sharp uptick in GAI adoption in industries such as healthcare \cite{hastings2024preventing} and policy-making \cite{huer2003challenges}, with the promise of unlocking unprecedented capabilities with far-reaching implications. However, GAI tools have also been found to embed biases that skew their generated artefacts towards the experiences of a select few rather than the diverse many \cite{septiandri2023weird, gupta2023sociodemographic}, prompting further scrutiny of their potential effects on society.

One such pattern of bias is a `Western gaze,' whereby outputs around non-Western contexts contain superimposed Western interpretations of those contexts \cite{kotliar2020data}. This is a critical problem, especially given how T2Is are being used globally and beyond just the West, and perhaps most prominently in Global South contexts such as Asia and Africa \cite{bianchi2023easily, gautam2024melting}. As representational systems \cite{hall1997representation}, T2Is thus have the power to make meaning and influence culture through produced artefacts and, by applying a Western gaze to Non-Western contexts, cause `representational harms' \cite{barocas2017problem} by misrepresenting and marginalizing the experiences of populations that are globally minoritized \cite{bender2021dangers}. However, this remains an under-addressed issue, as research into T2I ethics and fairness often center Western perspectives and rely on Western frameworks of ethics and fairness \cite{septiandri2023weird, das2024colonial, venkit2023nationality}.

We address these gaps through a Non-Western community centered investigation into depictions of non-Western cultures within T2I outputs, focusing on a specific context where AI-mediated harms are rising \cite{nyt_raj}: \textit{India}, an incredibly diverse country with over 1.4 billion people across many regional subcultures, each with their own cuisines, attire, art forms, festivals, and much more \cite{panda2004mapping}. Through grounded theory analysis \cite{charmaz2017constructivist} of 5 focus groups with 25 participants from various Indian subcultures/regions, we explore the following questions:  

\begin{itemize}
\item[] \textbf{R1:} How do T2I outputs represent Indian culture and its subcultures, and what are the implications of these representations for diverse cultural groups?
\item[] \textbf{R2:} What novel forms of representational harms emerge from the interactions between T2I systems and Indian cultural contexts, and how can we address these harms?
\end{itemize}

Beyond focus group participants perceiving representational harms defined by \citet{dev2020measuring} such as stereotyping of Indianness, the erasure and disparagement of specific subcultures and a lower quality of service in comparison to Western contexts  we document two novel cultural harms: 

\begin{enumerate}
    \item We present the novel representational harm of \textit{exoticism}, defined as ``the overamplification or overrepresentation of specific features or qualities of a culture in broad depictions of that culture, often at the cost of culturally-accurate details." Across 5 focus groups, we document exoticized portrayals of Indian culture through themes such as overrepresentation of rural settings and backdrops, female-presenting individuals being exclusively clad in traditional Indian sarees, and overtly colorful depictions of commonplace depictions of markets or everyday attire. We also find exoticist portrayals to be persistent, as participants' best efforts to engineer prompts to avoid such depictions of Indian culture bore no fruit. 
    \item We also present the novel representational harm of \textit{cultural misappropriation} within the outputs of T2Is, defined as ``the depictions of details about a culturally-specific context which incorrectly embeds details from one culture or subculture outside of its appropriate context." Some examples of cultural misappropriation observed within our focus groups include homogenizing commonly occurring Indian breakfast items from different parts of India onto the same plate, incorrectly presenting region-specific styles of wearing Indian clothing, and showing inaccurate ornaments for dance forms with centuries of historical and cultural contexts. In documenting this representational harm, we uncover how T2Is embed very surface-level understanding of details about Indian culture and its various subcultures.
\end{enumerate}

Having documented such harms perceived by individuals from a wide range of Indian communities within live-generated outputs of the T2I Stable Diffusion in response to prompts of their own choosing, we present guidelines for design principles that can inform the development of more inclusive and culturally sensitive T2Is that can accurately represent nuanced cultural interactions and experiences. We adopt a sociotechnical lens \cite{shelby2022sociotechnical}, recognizing the importance of culture-informed data work in alleviating these issues \cite{sambasivan2021everyone}. We contribute to the large conversation around designing safer T2Is in non-Western contexts and the Global South \cite[e.g.,][]{gautam2024melting, qadri2023ai, sambasivan2021everyone}, as we consider the application of our proposed design principles in other non-Western cultures not centered in T2I design and suffering similar representational harms.  

\section{Related Work}

\subsection{Biases and Harms within GAI Tools: \\ \hspace{2.2em}A Sociotechnical Systems Perspective}

The extensive documentation and operationalization of biases embedded within GAI tools to produce harmful outcomes has sparked widespread concern among researchers, who have highlighted the far-reaching negative consequences of these models on diverse communities \cite{bianchi2023easily, bender2021dangers, gupta2023calm}. As such outputs are seen to amplify societal biases that disproportionately affect marginalized groups by perpetuating a variety of racial \cite{field2021survey}, gender \cite{bolukbasi2016man, caliskan2017semantics}, disability \cite{gadiraju2023wouldn, venkit2023automated}, and nationality-based biases \cite{venkit2023nationality, narayanan2023unmasking} that reinforce outdated attitudes and exacerbating harmful stereotypes. 

Through the propagation of such stereotypes and the presence of negative biases, the outputs of GAI tools can and do cause harm \cite{das2024colonial}. In the broader context of machine learning-based systems, \citet{blodgett2021sociolinguistically} break these down into representational and allocational harm, based on whether harmful representations are generalized or resources are distributed disparately. A taxonomy of fine-grained representational harm categories includes quality of service, stereotyping, stigmatization, alienation, and public participation \cite{blodgett2021sociolinguistically}. Documenting attributes of bias measures can facilitate a better understanding of harms, use cases, and limitations. Building on this, \citet{dev2022measures} developed a framework categorizing harm into five types: \textit{Stereotyping, Disparagement, Dehumanization, Erasure,} and \textit{Quality of Service}. Their work offers a framework to GAI researchers such as ourselves for labeling and investigating patterns of harms. 

Investigating such harms has become especially critical as GAI systems are growing into a prominent part of the suite of AI-as-a-Service (AIaaS) tools that aim to provide convenient plug-and-play AI services \cite{lewicki2023out} across broad domains such as education \cite{rane2023contribution}, healthcare \cite{zack2023coding, hastings2024preventing}, and policy-making \cite{huer2003challenges}. The ease of using AIaaS tools has led to its meteoric rise in adoption in global contexts, and a subsequent surge in investigating how biases in such contexts propagate globally \cite[e.g.,][]{lewicki2023out, narayanan2023towards}. While designers of GAI and AIaaS tools have been working on mitigating such biases, the majority of approaches have been found to be mostly technical \cite{birhane2021algorithmic}, completely ignoring the social contexts in which systems operate \cite{venkit2023sentiment, venkit2024confidently}. 

Researchers have therefore argued that mitigation of biases should adopt a \textit{sociotechnical} approach, recognizing that AIaaS tools both influence and are impacted by end-users \cite{cooper1971sociotechnical, brynjolfsson2023generative}. This is critical since purely technical approaches miss the important fact that the data driving GAI tools embedded within AIaaS services is heavily influenced by the perspectives and positionalities of the humans who were part of model training teams \cite{birhane2021algorithmic}, and that such an oversight can lead to perceived technical `solutions' being ineffective in the real world \cite{d2020data}. A sociotechnical systems perspective on AI services and embedded GAI tools requires the perspectives of communities whose members experience the harms and suffer their impact in their daily lives. This approach prioritizes the equal involvement of social stakeholders and minority voices in the design, development and deployment of these sociotechnical systems, ensuring that their perspectives and experiences are not erased or marginalized. Community-centered approaches towards recognizing and mitigating the biases within GAI tools, such as the work of \citet{qadri2023ai}, \citet{mack2024they}, and \citet{prabhakaran2022cultural}, are thus essential. 

However, most research in understanding and mitigating harms has largely ignored such harms within non-Western contexts, with a few notable exceptions \cite{das2024colonial, qadri2023ai}. Our study aims to address this gap by adopting a community-based approach to investigate and define harms generated by AI models, particularly emphasizing cultural and subcultural representation within diverse communities across India.

\subsection{Stereotypical Representations of \\\hspace{2.2em}Indian Culture/ Subcultures}

Though depictions of India in Western media as `the land of snake charmers' \cite{chaudhuri2009snake} and other features which Satyajit \citet{ray1988} called `the false-exotic' might sound like relics of the past, that is not the case. As recently as the 2010 Julia Roberts classic `Eat, Pray, Love,' the protagonist perpetuates another famous Indian stereotype, as she visits India to `find herself.' Hollywood projections of India are littered with stereotypes, both Orientalist perspectives \cite{said1977orientalism} that show India as the “land of miracles and wonders,” \cite{chakkarath2010stereotypes} as well as ones about subcultures such as ``the artistic Bengali, the simple Gujarati, the austere Maharashtrian, and the intelligent Tamil brahmin" \cite{mehta2011bollywood}. Furthermore, given how films made in Hollywood or European film industries often are centered in their own contexts, Indian representation typically happens through characters in the Indian diaspora, either as fully Westernized individuals who ignore or actively look down upon elements of Indian culture (most commonly, eating with one's hands), or projecting a romanticized view of India as a place they long to return to \cite{trivedi2008bollywood}. While portrayals of Indian culture within Western film media might evoke a sense of India having established a presence on the global stage \cite{matusitz2012globalisation}, stereotypical representations of Indianness still leave much to be desired. 

However, any discussion of representations of Indian culture/subcultures is incomplete without addressing how they occur \textit{within} Indian contexts. Mainstream media within India, primarily dominated by India's Bollywood film industry, also perpetuates stereotypes around Indian culture/subcultures. As \citet{arora2023just} writes, ``popular Hindi cinema has articulated Indian identity as Hindu identity, with the normative Indian citizen presented as North Indian and Hindu." Bollywood movies across the past 70 years has overwhelmingly featured characters with traditionally Hindu names, with little to no geographical representation of the Northeastern states of India \cite{khadilkar2022gender}. Muslim characters, when shown, are often cast as violent antagonists or terrorists associated with Pakistan \cite{bhat2019muslim,kumar2013constructing}, or the `good Muslim' \cite{mamdani2002good} who stands against other antagonistic Muslims \cite{islam2007imagining}. Similarly, Northeast India, when featured, is often romanticized as exotic landscapes, contrasting the slow and rural lifestyles here with the busy, urban cities elsewhere in India \cite{dowerah2017cinematic,hasan2011talking}. These depictions of India contribute strongly to how Indian culture is perceived, defining `Indianness' to the West \cite{matusitz2011bollywood}.    

Our paper directly follows \citet{qadri2023ai}'s investigation into T2I representations of the Southeast Asian countries of India, Bangladesh and Pakistan. We extend this work by focusing specifically on Indian culture/subcultures, paying respect to their uniqueness within Southeast Asia and going against any suggestions of a regional monolith, and dive deeper into showing how the depictions of cultures/subcultures actively cause harms to individuals who identify with such cultures/subcultures. 

\section{Methods}
\subsection{Recruitment and Focus Group Descriptions}

To analyze how Indian culture and various subcultures are represented within generative AI systems, we conduct a set of community-centric focus groups with participants from these cultures. We chose focus groups instead of individual interviews to both mirror \citet{qadri2023ai}'s study towards similar goals, and because such a method is known to be effective in empirically studying cultural contexts \cite{hughes1993using,rodriguez2011culturally} where participants can collectively negotiate shared understanding of what `culture' means to them \cite{huer2003challenges}. 

We recruited participants through advertising the study through our social networks and postings in public places, inviting participation from individuals above the age of 18 and self-identifying as being Indian by virtue of growing up in India or not having lived outside of India for over 10 years. We successfully recruited 25 participants. Among these, 60\% self-identified as female, while 40\% self-identified as male. We asked participants which of the five regions of India (Northern, Eastern, Western, Southern, or Central) they self-identified as being from. Our study represents 4 participants from North India, 6 from East India, 5 from West India, 9 from South India, and 1 from Central India. We randomly sorted participants into 5 focus groups of 5 individuals each, and not grouping participants from similar regions to avoid groupthink bias where like-minded participants would only have shallow conversations \cite{macdougall1997devil}, with each focus group being conducted Oct-Dec 2023 by two researchers in accordance with best practices \cite{halcomb2007literature}. We present participant distribution information in Table \ref{table:participant} (Appendix C), alongside assigned pseudonyms inspired by renowned literary figures from diverse cultural and subcultural backgrounds across India, to avoid numeric identifiers while still using culturally-appropriate names as opposed to Western pseudonyms commonly used. By using \textit{culturally meaningful pseudonyms}, we avoid performing cultural erasure while respecting their anonymity as we highlight quotes throughout this paper. 

Within each focus group, we began with providing participants with a survey soliciting examples of prompts for which they would like to see outputs from \textit{Stable Diffusion v2.1}: a T2I which takes in text-based prompts to generate images. This version was selected due to its open-source nature, providing free access and flexibility to be used across various platforms with evidence and ability of usage as a sociotechnical system across various social spaces. Once participants finished providing prompts in their surveys, we asked icebreaker questions around their usecases and general opinions around GAI tools, while researchers analyzed provided prompts for similarities and themes to determine the most common prompts across all responses. Having obtained the prompts, we generated Stable Diffusion outputs live in front of participants, so as to be transparent around how we were obtaining outputs. Each output was a 2x2 grid of four images per prompt, a number chosen to balance the amount of time it takes to generate a live image with the intent of showing participants a variety of outputs while showcasing the standard output syntax provided by T2I models. We displayed each output set of images to participants, one at a time, and asked them an open question around their thoughts on the images to hopefully hear their most honest thoughts. 

We asked participants to reflect on specific examples on areas they believed the images were to their satisfaction, as well as where they were dissatisfied, inviting them to discuss within the group rather than us asking each participant individually. We also offered participants the option of providing us with any `follow-up' prompts for which they would like to see the outputs to, if such outputs could be used in comparison with those from the original prompts. Finally, we asked participants whether they believed that a given prompt output was an image they would show to non-Indian individuals as good representations of Indian culture or one or more subcultures, and why/why not. In support of open science and ensuring transparency in our research process, we have made the anonymized materials used for this study available in our repository\footnote{\url{https://github.com/PranavNV/Harms_In_GAI}}. This includes quotes from focus groups, coding frameworks, recruitment survey, generated images, and other data that can contribute to a better understanding of our work. The study received Exempt status by Institutional Review Board (IRB) review.

\subsection{Thematic Analysis}\label{subsec:analysis}
We analyzed our focus group data through a constructivist grounded theory approach \cite{charmaz2017constructivist}, individually coding transcripts line-by-line in accordance with \citet{charmaz2006constructing} to generate theories through inductive reasoning \cite{glaser1967discovery,glaser1992basics}. We conducted qualitative coding on the platform Taguette\footnote{\url{https://app.taguette.org/}}, creating codes and associated definitions for transcript snippets, and later synthesizing them through discussion within the research team.

\section{Findings}

In this section, we document our findings from the 5 focus groups, as we explicate the representational harms they observed with respect to Indian culture/subcultures. We note that 68\% participants (17/25) indicated using GAI tools multiple times per week and a further 20\% (5/25) using it at least once a month, underscoring participants' high familiarity with GAI tools. Participant usage of GAI tools was also in a wide range of cases, covering both text-based tools for tasks like writing support and gaining new knowledge, as well as T2Is for creating artistic content for both professional and personal purposes. Such patterns of usage also led to participants having opinions about perceived weaknesses within GAI tools. Participants were aware of GAI tools exhibiting \textit{``repetitiveness, over-generalization, and misleading tendencies." - Chitra}, while also noting that \textit{``The training data and outputs are biased, leading to harmful content aimed at marginalized groups." - Sonali}. Such comments depicted them to be effective judges of potential representational harms within GAI tools, particularly around cultural/subcultural contexts they identified with. 

\subsection{Exoticism within the Outputs of T2Is}\label{subsec:exotic}

\begin{figure*}[t]
\centering
\begin{subfigure}[t]{0.2\textwidth}
    \fbox{\includegraphics[width=\textwidth]{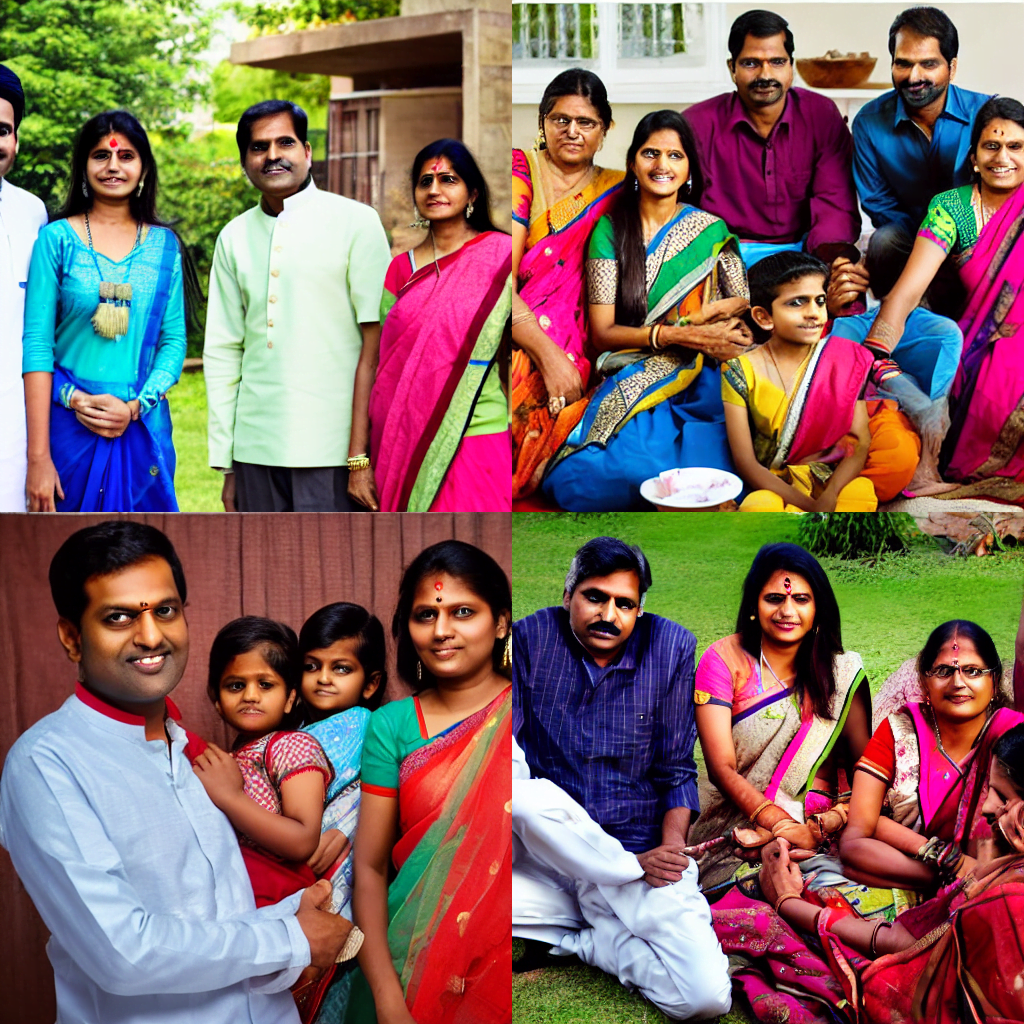}}
    \caption{T2I outputs for \\`Middle class Indian family'}
    \label{fig:middle-class}
\end{subfigure}
\hspace{0.5em}
\begin{subfigure}[t]{0.2\textwidth}
    \fbox{\includegraphics[width=\textwidth]{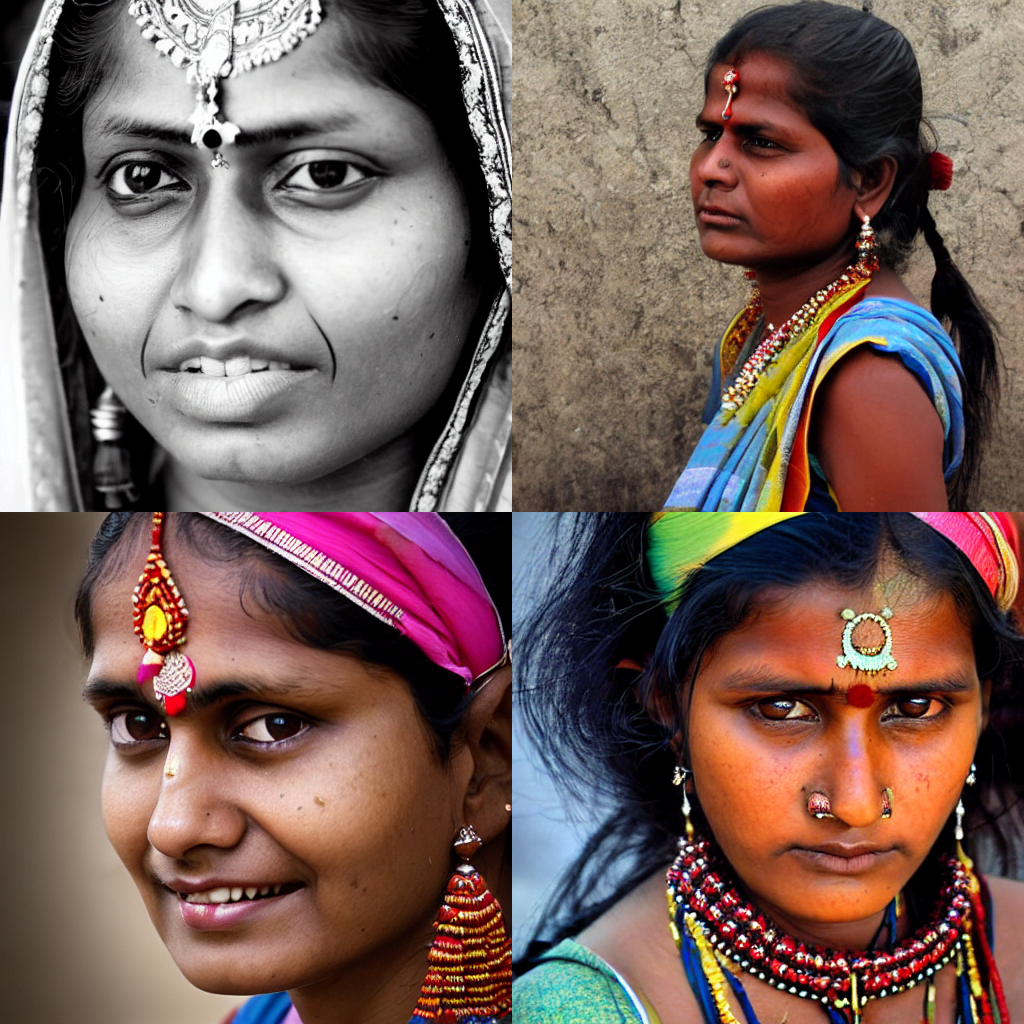}}
    \caption{T2I outputs for\\ `Indian woman from South Bombay'}
    \label{fig:south-bombay}
\end{subfigure}
\hspace{0.5em}
\begin{subfigure}[t]{0.2\textwidth}
    \fbox{\includegraphics[width=\textwidth]{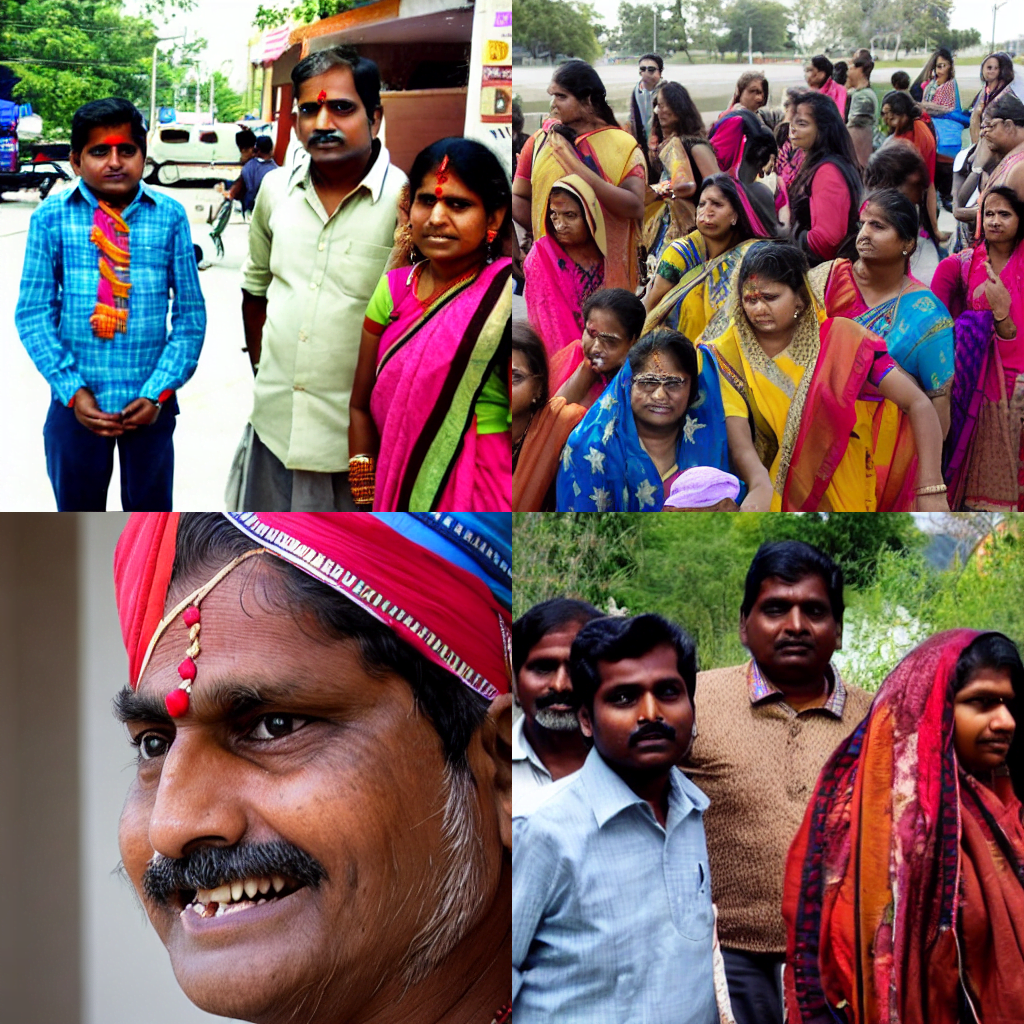}}
    \caption{T2I outputs for\\ `Indian people in America'}
    \label{fig:indian-america}
\end{subfigure}
\hspace{0.5em}
\begin{subfigure}[t]{0.2\textwidth}
    \fbox{\includegraphics[width=\textwidth]{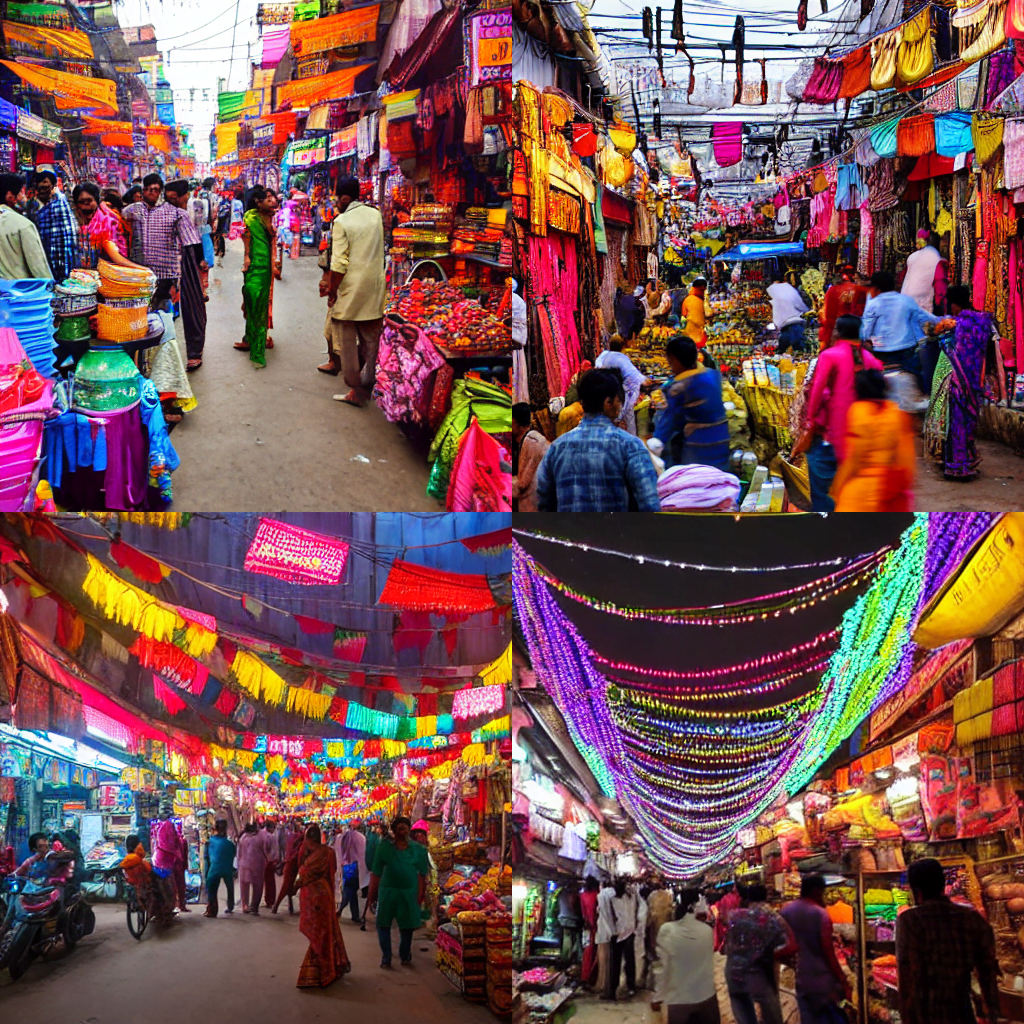}}
    \caption{T2I outputs for \\ `Indian market'}
    \label{fig:market}
\end{subfigure}
\caption{2x2 T2I Outputs showing Exoticized depictions of Indian culture}
\label{fig:exotic}
\end{figure*}

A primary theme emerging from our focus groups was a novel type of harm within the outputs of T2Is which we label \textit{exoticism}. While exoticized portrayals of India as `distanced' from Western traditions \cite{sen1997indian} under the larger umbrella of Orientalized perspectives of Asia within Western media \cite{said1977orientalism} are well-established, we document how members of Indian communities perceive T2I tools exacerbating and perpetuating such portrayals through their outputs. We thus define exoticism within the outputs of T2I tools as ``the overamplification or overrepresentation of specific features or qualities of a culture in broad depictions of that culture, often at the cost of culturally-accurate details."

A prominent way in which exoticism was present within the outputs of T2Is was in the depicted outfits of female-presenting Indians, as shown in the output for the prompt `Middle-class Indian family' in Figure \ref{fig:middle-class}. Aside from noting the strong heteronormativity present within all four image results within this output appearing to depict a male-presenting individual with a female-presenting individual, our participants also observed how the female-presenting individuals all appeared to be wearing sarees: a traditional Indian garment commonly worn within femme-identifying individuals. Participants were troubled with the implication that \textit{``just because [they] said `Indian', they're showing sarees? Because generally, middle class families don't always wear sarees'' - Arundhati}, a concern which is compounded by the fact \textit{``many men [are] wearing Western clothes, but not a single woman wearing them" - Mamta}. Such a strong prevalence of sarees is a common example of how the West perceives Indian women through their sarees \cite{benedict2024deconstructing, arora2023just}.

This depiction of Indian women in ``traditional" attire seemed impossible to shake off, as participants provided different prompts but the outputs kept showing sarees. For instance, the outputs for the prompt `Indian people in Western attire' (Figure \ref{fig:western}, Appendix \ref{appendix-images}) still overwhelmingly depicted women in sarees, leading Arundhati to wonder: \textit{``where is the Western attire??"} While outputs for the prompt `Indian people in Suits' (Figure \ref{fig:suits}, Appendix \ref{appendix-images}) showed suit jackets, it almost exclusively depicted male-presenting individuals and when participants expressly asked for `Indian women in Suits' (Figure \ref{fig:women-suits}), they were once again entirely shown sarees. Kamala noted that the model's reliance on majority datasets leads to the perpetuation of harmful notions, stating, \textit{``The model is showing these kinds of images because it's trained on majority datasets, where majority of the people and majority of the information that's fed into it is Indian people in their sarees. That's a stereotype, and that's why it's giving these kinds of images''} when the model chose only to depict women in sarees even when explicitly prompted to not do so by mentioning the desired attire. We thus document a pattern of the strong exoticism of Indian women within T2I outputs, as these outputs exclusively portray them as wearing sarees and refuse to present a different attire. This exoticism was deeply frustrating to participants, a sentiment best encompassed by Chitra: \textit{"Oh my God! This is so demeaning. Do they not have the concept of women wearing shirts and pants at all?"}

The strong refusal within T2I outputs to deviate from an exoticized stereotype about Indian culture is also evident when examining the overtly impoverished depictions of India. This began when outputs for `Indian woman' (Figure \ref{fig:indian-woman}, Appendix \ref{appendix-images}) showed what participants such as Amit identified as \textit{``village women"}, although participants were not too perturbed by this given that almost 65\% of India's population was rural at the last time of counting \cite{indiacensus}. However, similarly as above, participants were most offended when the model outputs refused to deviate from visuals about rural India. Such a deviation was particularly prominent within the outputs of the prompt `Indian woman from South Bombay' (shown in Figure \ref{fig:south-bombay}), which continued to show a very similar slate of images and characteristics as in the output for `Indian woman' (Figure \ref{fig:indian-woman}, Appendix \ref{appendix-images}). It was jarring for our participants to see that the addition of information within the prompt did nothing more than provide \textit{``just an iteration of results" - Anita}. Furthermore, the specific location added to the prompt is also pertinent: South Bombay is one of the most affluent localities within India, with real estate and property prices being among the highest in the world \cite{Babar24}. Therefore, the output showing figures perceived to be impoverished was difficult for participants to reconcile. Furthermore, even mentions of non-Indian locations were ignored within prompt outputs, as seen in the outputs for `Indian person in America' (Figure \ref{fig:indian-america}). Focus group participants noted how the outputs \textit{``totally ignored the keyword America" - Kamala}, and continued to show grasslands and architecture that they identified to be familiar within their Indian contexts. While rural and impoverished depictions of India have previously been explored by \citet{qadri2023ai}, GAI outputs refusing to change that even based on reprompting is a novel finding. 

Another pattern of the exoticized depictions of Indian culture was uncovered by T2I model outputs which contained an overabundance of bright and vibrant colors when depicting commonplace visuals. Although participants were aware of how \textit{``India has been, in the Western lens, always shown as a very colorful country," - Jhumpa}, they were nevertheless surprised to see just how exaggeratedly colorful depictions of Indian cultures were. This was particularly salient in outputs for the prompt `Indian market' (Figure \ref{fig:market}), which shows bright splashes of pink, neon blues and greens, and orange stringlike objects hanging over streets lined with vendors. Participants such as Namita noted that while the depictions of the vendor stalls seemed mostly plausible, the image was soured by being \textit{``too colorful."} Sonali also elaborated that while they have seen markets be colorfully decorated, especially in the evenings, \textit{``but not as colorful as these images are."} Furthermore, depictions of Indian festivals (Figure \ref{fig:festival}) also showed individuals dressed in colorful clothes or wearing colorful garlands of flowers. While such attires are not uncommon in some festivals in India, they certainly do not represent all festivals, such as prominent festivals in Southern India for which \textit{``it's very rare that there's so many bright colors, and it's usually very muted" - Surya Kumar}. Indeed, as Jhumpa put it: \textit{``I think it does stereotype the whole idea of `oh, India's a very colorful country, everything has a lot of color, and all of our clothes are shiny and colorful,' which I'm not a huge fan of."}

Thus, we document strong patterns of exoticism within T2I outputs, where the outputs persistently overrepresent specific stereotypes associated with India, and resist prompt-engineering attempts to avoid them. 

\begin{figure*}[t]
\centering
\centering
\begin{subfigure}[t]{0.2\textwidth}
    \fbox{\includegraphics[width=\textwidth]{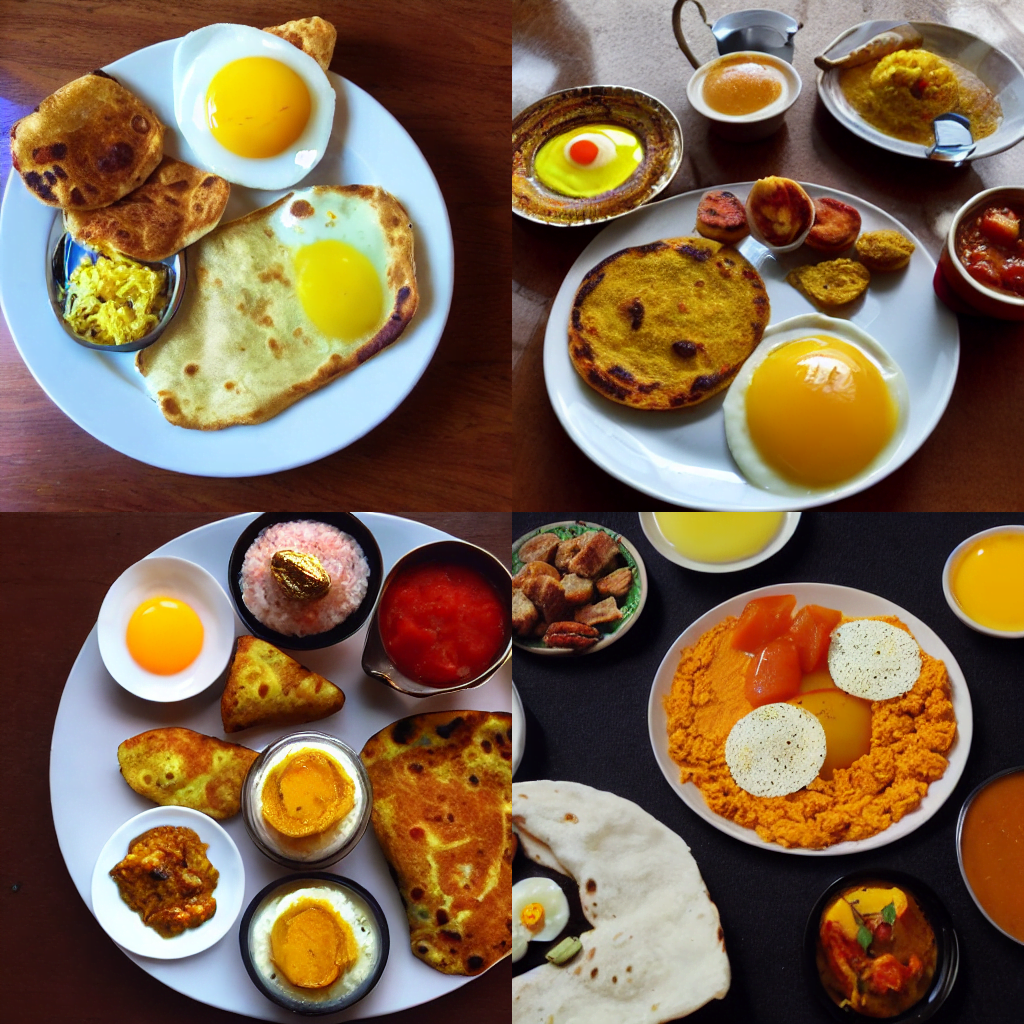}}
    \caption{T2I outputs for\\ `Indian breakfast'}
    \label{fig:breakfast}
\end{subfigure}
\hspace{0.5em}
\begin{subfigure}[t]{0.2\textwidth}
    \fbox{\includegraphics[width=\textwidth]{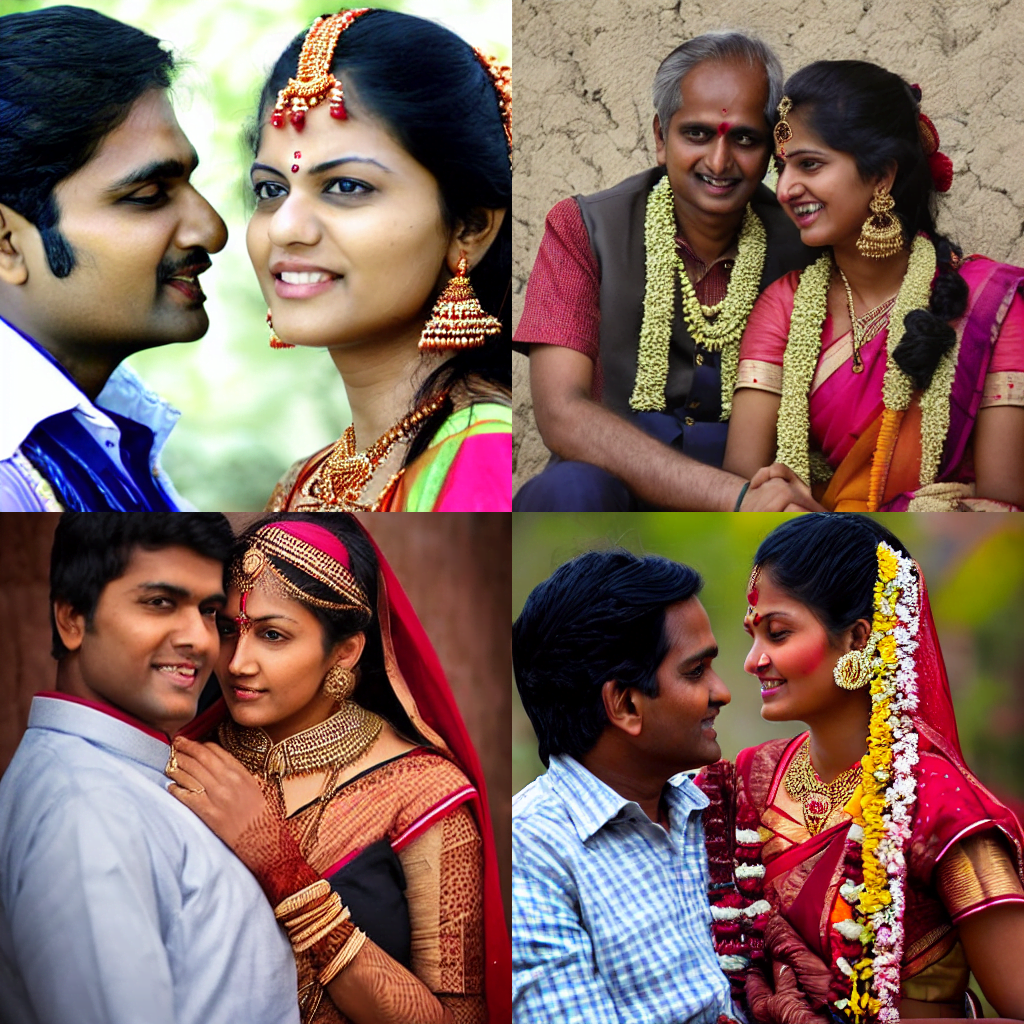}}
    \caption{T2I outputs for \\`Indian couple'}
    \label{fig:couple}
\end{subfigure}
\hspace{0.5em}
\begin{subfigure}[t]{0.2\textwidth}
    \fbox{\includegraphics[width=\textwidth]{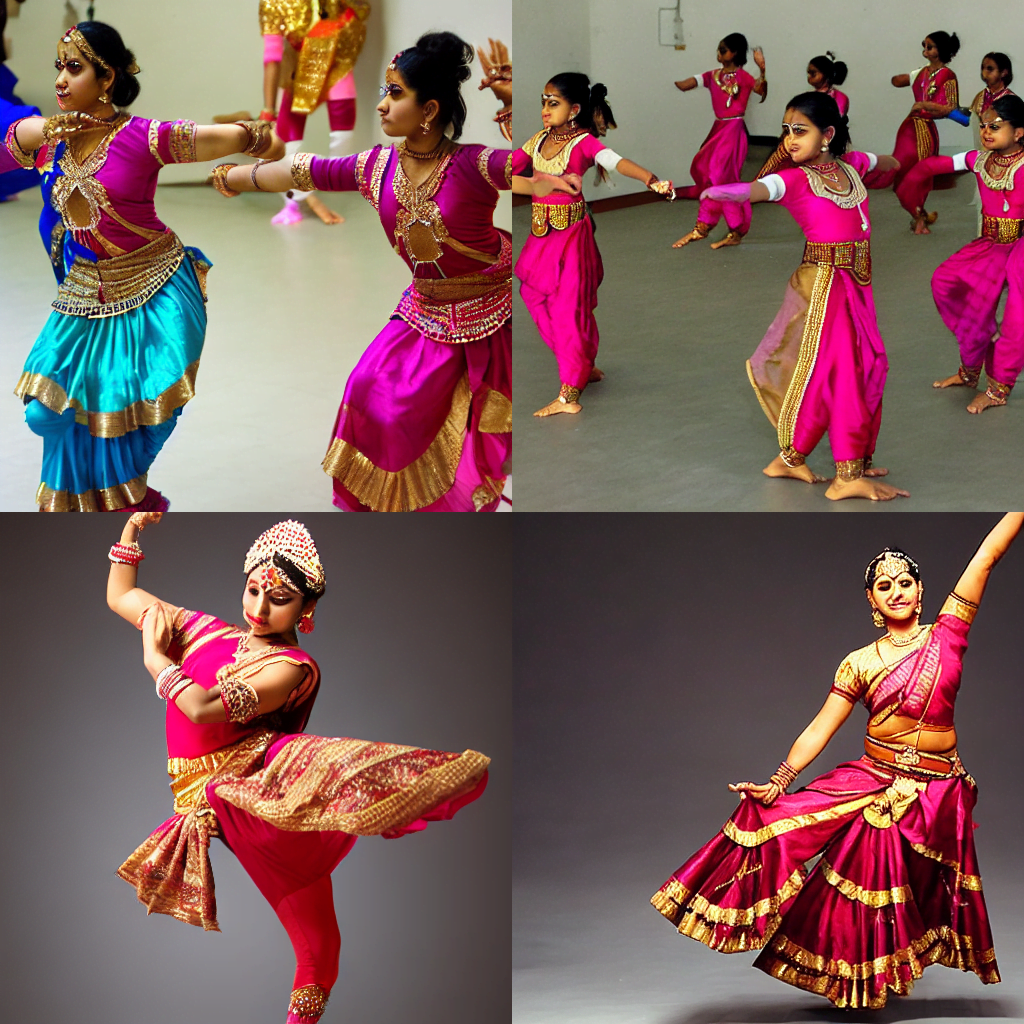}}
    \caption{T2I outputs for\\ `Indian dance'}
    \label{fig:dance}
\end{subfigure}
\hspace{0.5em}
\begin{subfigure}[t]{0.17\textwidth}
    \fbox{\includegraphics[width=\textwidth]{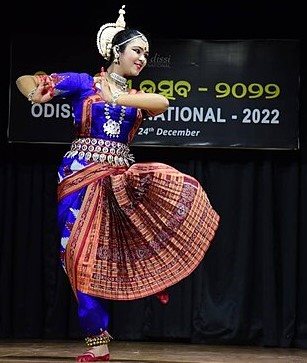}}
    \caption{Traditional Odissi Dancer, from Wikimedia Commons}
    \label{fig:odissi}
\end{subfigure}
\caption{2x2 T2I Outputs showing Culturally Misappropriated depictions of Indian culture}
\label{fig:exotic1}
\end{figure*}

\subsection{Cultural Misappropriation within the Outputs of T2Is}

Through our analysis of participant responses, we observe the emergence of a novel type of harm within the outputs of T2Is: \textit{cultural misappropriation}. While the concept of `misappropriation' is most commonly used in the context of external forces such as colonial or imperial powers taking or using aspects of a culture to which they do not belong, we define cultural misappropriation within the outputs of T2Is as ``the depiction of details about a culturally-specific context which incorrectly embeds details from one culture or subculture outside of its appropriate context." 

One of the most prominent ways in which the outputs of T2Is cultural misappropriation is in the homogenization of components from multiple cultures into a single rendition. This was most prominent in the Stable Diffusion results for the prompt `Indian Breakfast', shown in Figure \ref{fig:breakfast}. Focus group participants who viewed this image noted how \textit{``it put everything under one single bracket" - Kamala}, featuring components of breakfasts from various parts of India, rather than each individual image featuring a plate from a single region. Participants mentioned expecting to see culturally-accurate plates from different parts of India, especially since they were informed that they would be shown 4 results for the same prompt. Instead, participants such as Chitra were dissatisfied at such homogenized images being \textit{``clearly a bad mix of regions"} that did not respect the uniqueness of any regions. Anita, a vegetarian from the primarily-vegetarian Indian state of Gujarat, particularly took offense at what they perceived to be sunny side-up eggs present in each image, which did not respect how sizeable sections of Indian societies are strict vegetarian and not the assumed ovo-lacto vegetarians. Thus emerged the pattern of Stable Diffusion homogenizing results from different subcultural contexts into a single depiction, and thus flattening a diverse range of results into a single notion of Indian-ness.

Additionally, Stable Diffusion outputs also demonstrated cultural appropriation by depicting details in images that was not specified in prompts. A strong example of this is around the outputs of the prompt `Indian Couple', which is shown in Figure \ref{fig:couple}. These images contain the same prevalence of heteronormative couples with the female-presenting individuals clad in sarees as discussed in Section \ref{subsec:exotic}. However, sarees are made of several different types of fabric with a multitude of weaving styles, and are worn differently in different regions of India. Looking at the image outputs, participants noted how the depicted individuals looked more like couples on their wedding day, as made evident by the extravagance of the sarees and jewelry they were adorned with. Participants like Birinchi also remarked that while depicted individuals looked South Indian, the sarees they were shown in appeared to be closer to those prevalent in North India. This was a common observation across our focus groups, as Chitra and Namita also noted similar mismatches within outputs where female-presenting individuals were depicted as clad in sarees that were incongruous to the region of India they were perceived to be from e.g., an individual perceived to be a woman from Northern India shown wearing ornaments from Northern India but a saree from Southern India. It is not just about the region where sarees originate and are worn by individuals from a different region, which is plausible through tourism, but the styles of wearing sarees are also specific to different regions of India. Within our generated Stable Diffusion outputs are documented several instances of such mismatches, such as female-presenting individuals wearing what are perceived to be `Kanjivaram' sarees from Southern India in styles that are more prevalent in Northern India. This is a significant problem, since such mismatches are deeply disrespectful to the saree and the traditions and histories of labor and culture it stands for \cite{deepshikha2018textiles}. 

Another instance of cultural misappropriation within Stable Diffusion outputs was in response to the prompt `Indian dance', shown in Figure \ref{fig:dance}. The outputs show various figures, all appearing to be performing what, based on the angles of the dancers' bodies and arms, participants such as Mamta stated \textit{``I think [these are] Odissi"}. However, the composition of the images demonstrated what participants considered a strong cultural inaccuracy: that of ornaments. Indian classical dances such as Odissi are steeped in centuries of history and tradition, with styles of movement, attire, makeup, and ornaments unique to each dance form. In the depicted images, while the dance forms is primarily identified as Odissi based on the dress and body positions, a glaring culturally inaccurate feature is in the headgear ornaments. Odissi is characterized by headgear (see Figure \ref{fig:odissi}) which traditionally is near the back of a dancer's head \cite{pattnaik2017comparative}, whereas the depicted images all show jewelry towards the front and above the forehead. What participants such as Jhumpa recognize within the depicted images as \textit{``temple dances, [because of] the gold and the gold borders} are perhaps features closer to the Southern Indian dance form of Bharatnaytam \cite{pattnaik2017comparative}, though the posture of the dancers seem to point towards the Eastern Indian dance form of Odissi. As Vikram remarked, \textit{``it looks like it's trying to emulate generic stuff that Indian women put on their heads. But it has lost that specificity, and I can't place what these ornaments are."} This confusion between different Indian classical dances, each a product of rich cultural lineages, is a strong example of cultural misappropriation within T2I outputs, that carelessly homogenize and mischaracterize elements of multiple subcultures, thus dishonoring the traditions they represent.

Thus, we observe several instances of T2I outputs containing cultural misappropriation, where components from several unique cultural components are homogenized into one representation, which ultimately ends up being inaccurate and offensive to all the subcultures from which it draws. 

\subsection{Documenting Representational Harms \\\hspace{2.2em}within T2I Outputs}

We also document evidence of participants identifying different types of representational harms as documented by \citet{dev2020measuring}: \textit{stereotyping, disparagement, dehumanization, erasure,} and \textit{quality of service}. We particularly focus on instances where more than one of these occur concurrently. 

\subsubsection{Stereotyping and Erasure}
\citet{dev2020measuring} defined the harm of stereotyping as `overgeneralized beliefs about the personal attributes of an individual as determined by their demographic group membership,' while erasure is `the lack of adequate representation of members of a particular social group.' They note how these typically have a `cyclical relation,' a phenomenon our participants also pointed to. 

One aspect along which participants, across multiple focus groups, noted the harms of stereotyping and erasure was religion, as Stable Diffusion outputs seemed to depict visuals commonly associated with Hinduism. Participants such as Satyajit, Arundhati, and Vikram noted that the outputs for `festivals of India' (Figure \ref{fig:festival}) featured the bright colors, flower-knit garlands, festive marks on depicted individuals' foreheads, and other attire and jewelry consistent with Hindu festivals. Missing in these representations was any evidence or visuals prominent in Islamic, Christian, Buddhist, or Sikh festivals, to name a few excluded religions. The erasure of other religions is also evident in outputs around `Indian weddings', which prominently feature Hindu marriage rituals and visuals such as ornaments and attire, and omit representations of Islamic wedding practices such as the `nikkah' or the church altars of Christian weddings.

The prominence of visuals associated with Hinduism is not entirely unexpected, since almost 80\% of India is Hindu \cite{indiacensus}. What is harmful is the fact that even within Hinduism, depictions seem to only feature Hindu festivals prominent in Northern India, which only accounts for about 28\% of India's population \cite{indiacensus}. Surya Kumar, a participant from the South Indian state of Kerala, noted that outputs did not \textit{``reflect any festival that we celebrate in the South."} Birinchi Kumar and Satyajit also noted the absence of other Hindu festivals such as Durga Puja or Chhath Puja celebrated in Eastern India by almost 200 million people, and Pongal celebrated in Southern India by almost 100 million people. Within the context of weddings, outputs prominently feature components such as `sangeet', predominantly seen in North Indian Hindu weddings and not as common within weddings in the South or East of India. This is best exemplified by Aravind: \textit{``it seems more like a traditional North Indian wedding rather than a South Indian wedding."} T2I outputs around Indian culture thus portray a strong stereotype of North Indian Hinduism being the default form of Indian-ness, erasing other Hinduisms and other religions practiced in India. 

The propagation of the default depiction of Indian-ness heavily skewing towards North Indian subcultures and erasing others is also prevalent across a range of other prompts. For instance, participants interacting with `Indian breakfast' (Figure \ref{fig:breakfast}) lamented the absence of food items and breakfast staples from Southern India, with Kamala noting that \textit{``I can’t see any dosa, or anything of that sort."} Similarly, in the context of Indian dances (Figure \ref{fig:dance}), participants such as Mamta noted how \textit{``certain classical dances from the Northeast, such as Sattriya or other dances, are not depicted in these images."} These results indicate a prevalence of a very specific type of Indian-ness within T2I outputs corresponding to North Indian subcultures, erasing any others. 

\subsubsection{Disparagement and Quality of Service} Disparagement is the `behavior which reinforces the notion that certain groups are less valuable than others and less deserving of respect,' which also affects quality of service `where a model fails to perform equitably for different groups.' \cite{dev2020measuring} The intersections of these harms emerged where participants observed outputs for similar prompts containing radical differences when a single word/phrase changed. 

\begin{figure*}[t]
\centering
\centering
\begin{subfigure}[t]{0.2\textwidth}
    \fbox{\includegraphics[width=\textwidth]{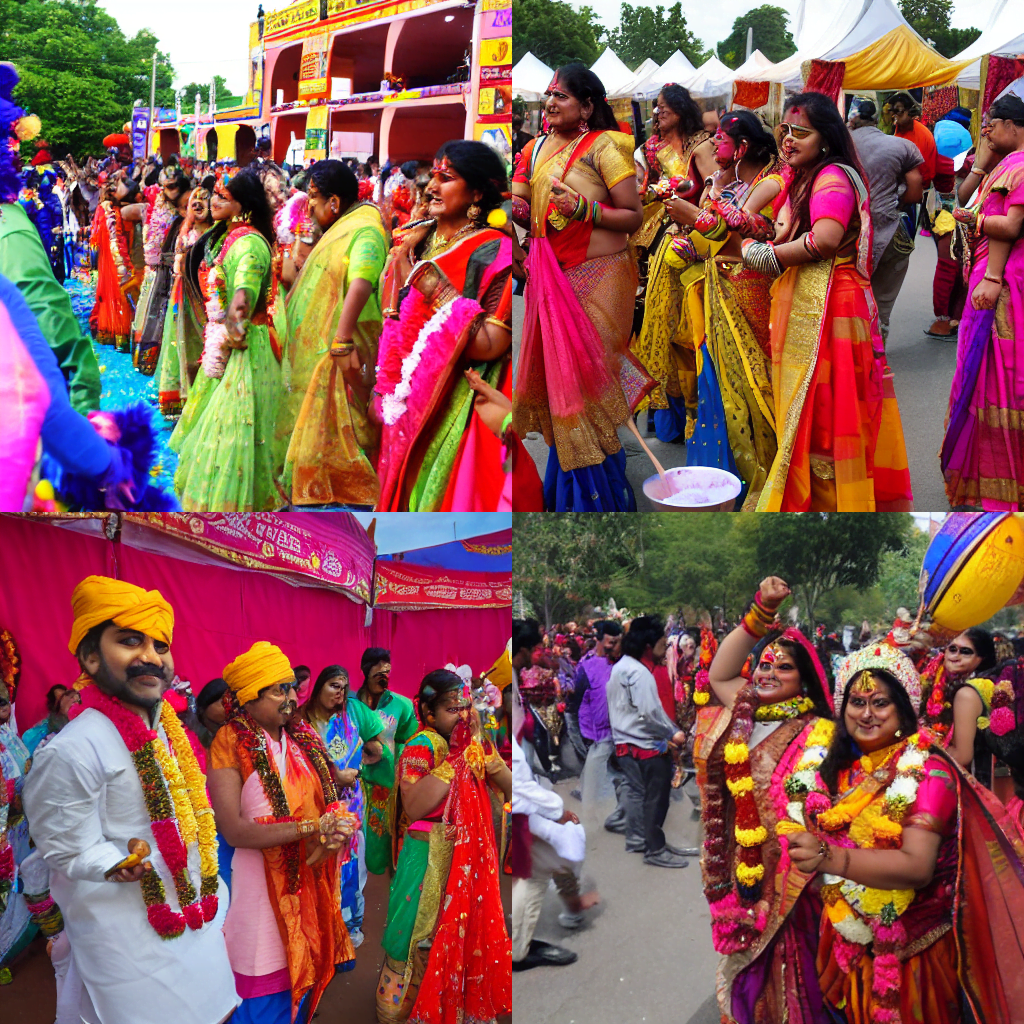}}
    \caption{T2I outputs for \\`Festivals of India'.}
    \label{fig:festival}
\end{subfigure}
\hspace{0.5em}
\begin{subfigure}[t]{0.2\textwidth}
    \fbox{\includegraphics[width=\textwidth]{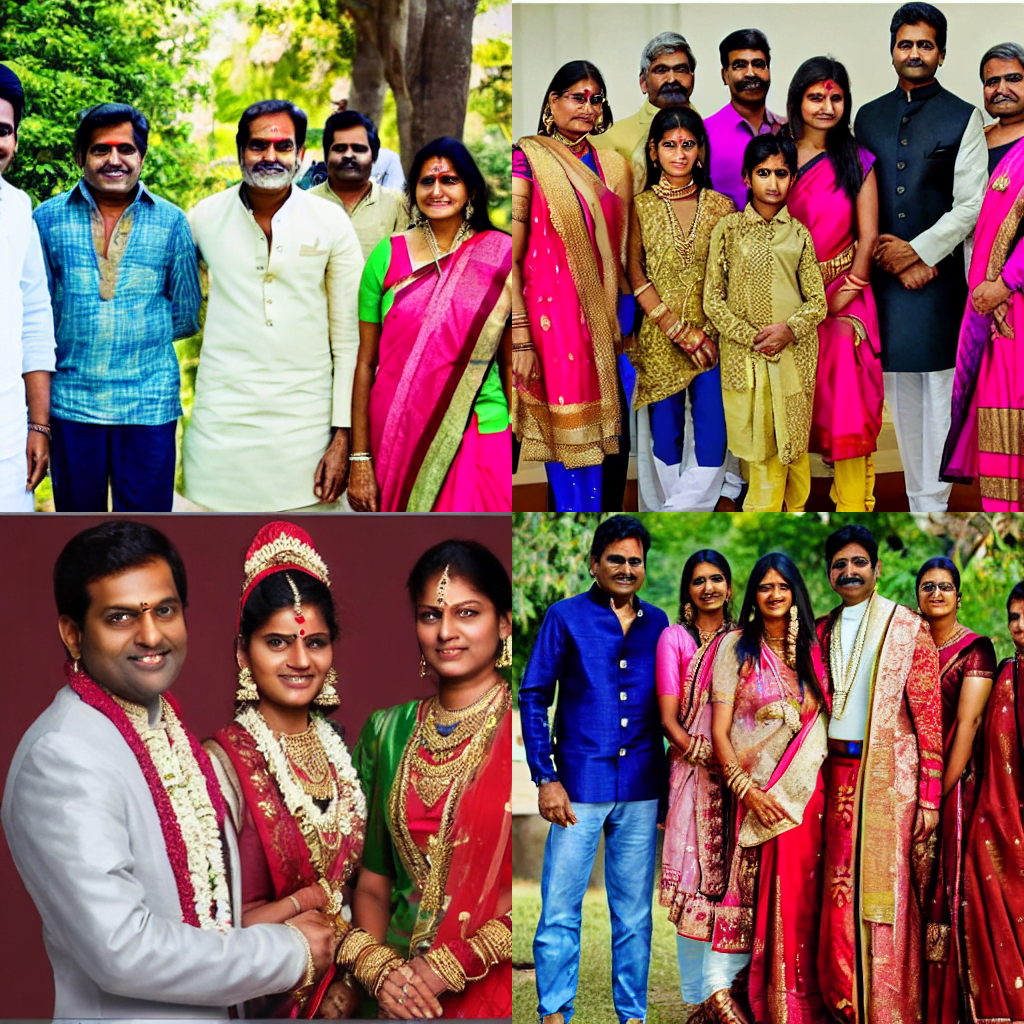}}
    \caption{T2I outputs for \\`Rich Indian family'}
    \label{fig:rich}
\end{subfigure}
\hspace{0.5em}
\begin{subfigure}[t]{0.2\textwidth}
    \fbox{\includegraphics[width=\textwidth]{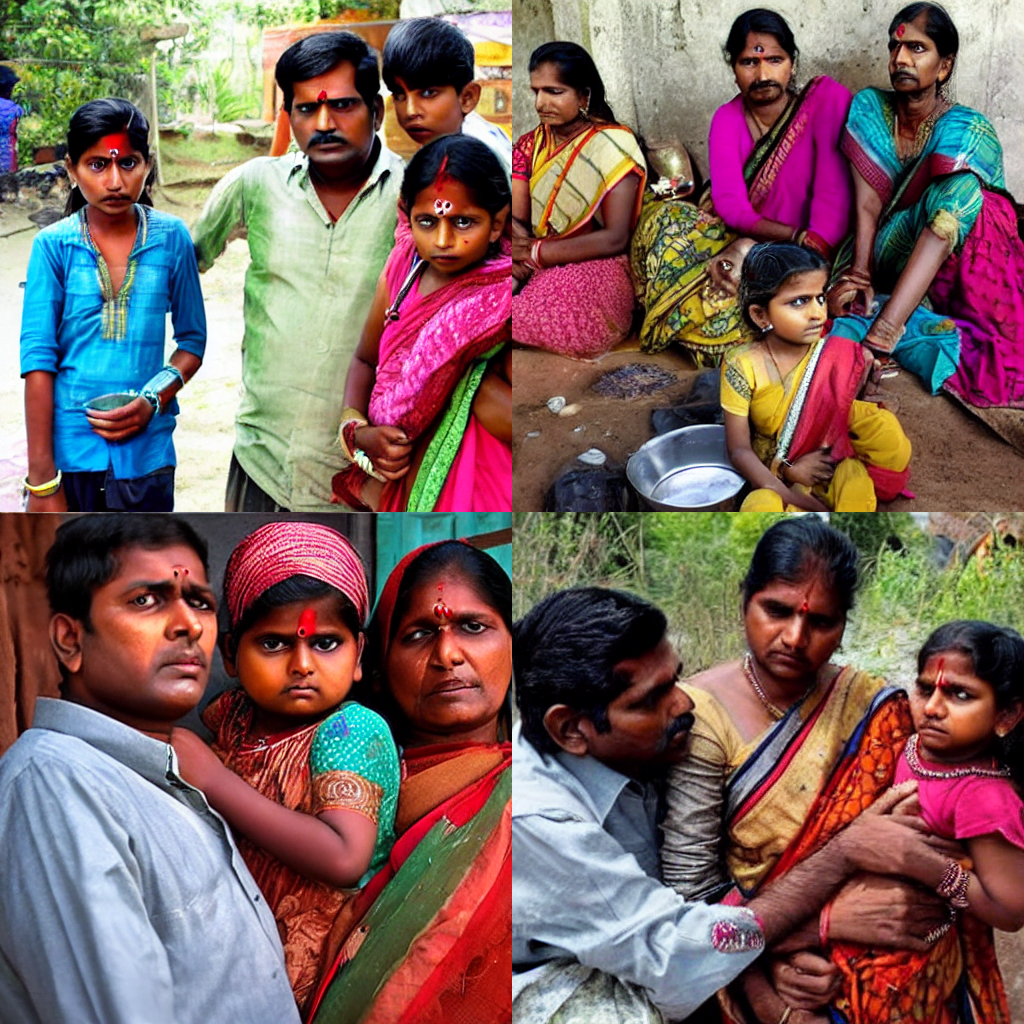}}
    \caption{T2I outputs for\\ `Poor Indian family'}
    \label{fig:poor}
\end{subfigure}
\caption{2x2 T2I Outputs showing Representational Harms within depictions of Indian culture}
\label{fig:five-harms}
\end{figure*}

When comparing the T2I outputs to the prompts `rich Indian family' (Figure \ref{fig:rich}) to those of `middle-class Indian family' (Figure \ref{fig:rich}) and `poor Indian family' (Figure \ref{fig:poor}), participants noted that while the faces of the depicted individuals were mostly identical across all cases, there were large differences in overall attire and background of images, with depictions of poverty being especially prominent as those of rich and middle-class families stayed close to each other. Chitra noted how \textit{``the amount of jewelry that they are wearing"} was different, with the rich and middle-class families being adorned with large golden ornaments whereas the poorer individuals were shown to wear smaller and duller ornaments, in addition to the rich and middle-class families wearing \textit{``sarees of better quality of cloth"}. The colorful nature of the attire and smiling faces present within the outputs of the rich and middle-class families stood out in stark contrast to the muted clothing and frowning faces assigned to the poorer families, leading Sudha to inquire, \textit{``Why are only the rich people celebrating things when the poor people are not celebrating something?"} Finally, Kamala also noted that \textit{``looking at their complexions, for poor they’re depicting it dark. I don't know if it's just me, or done like that to showcase that they're poor,"} alluding to colorist tendencies of equating poverty with darker skin \cite{peters2021colorism}. This indicates a clear pattern of T2I outputs disparaging poorer sections of Indian society as compared to rich or middle class families, offering a lower quality of service to over 232 million Indians classified as poor \cite{indiacensus}.

\subsubsection{Disparagement and Stereotyping} 
Several participants expressed concerns regarding the portrayal of their cultures, highlighting a prevalent reliance on stereotyped Western perspectives aligning with the disparagement of their own experience. Sudha noted, \textit{``I know the discussion is about India, but I think it's mostly a representation of how US and UK sees us.''} This is particularly problematic given the global reach of these models, as the dissemination of misaligned information can have far-reaching consequences.

The generated images often depicted India in stereotypical dichotomies, portraying it either as a colorful, tradition-rich country or as impoverished and tribal. Participants deemed these representations as reflective of an external viewpoint rather than an authentic portrayal of India, disparaging or not prioritizing their own experiences. Jhumpa noted, \textit{``India has been, in the Western lens, always shown as a very colorful country, and so when you make an illustration as opposed to the generator showing a real image, the color can be exaggerated.''}. This behavior was evident across several prompts, with Stable Diffusion consistently amplifying colors in depictions of Indian markets, attire, and festivities, deviating from their authentic representations and presenting a wrongful portrayal of culture. We thus see an accentuation of a Western gaze across our findings.  

These interactions were treated to be a stereotyped notion of what the various cultures in India are, highlighting the `Whiteness of AI' as explained by Chetan: \textit{``The images which have been fed to these models are through the lens of mostly a white person visiting India. And then these models are trained on top of that, and they’re just reiterating what a white person sees in India. This shows the Whiteness of AI.''} These interactions demonstrate how the model's majoritarian bias yields a distorted understanding, perpetuating cultural stereotypes while marginalizing and disparaging minority experiences and perspectives.

\subsubsection{Dehumanization and Erasure}

Participants' responses highlighted the issue of dehumanization where certain groups, particularly women and minorities, are marginalized in the outputs and treated `lesser than human'. For instance, Jhumpa pointed out in a comparison between outputs for `Indian God', \textit{``Female bodies are penalized as being more sexual, for much less. There were torsos shown in the other picture as well, but those were safe for work,"} 
as they noted that the model did not generate images of `Indian Goddess' as it was deemed to be NSFW but showing no such censorship for `Indian God'. Such outputs perpetuate a harmfully objectifying gaze, reinforcing societal biases around women. 

The model also erases the experiences of cultural and regional minorities and \textit{``super duper minorities,"} as Kamala put it, who are not represented in the generated images by either choosing not to provide the right context or overlaying them with `cultural majorities' viewpoint, such as seen with festivals, celebrations and cuisines. Arundhati echoed this concern, stating, ``\textit{You can't see diversity in these pictures, from different parts of India, like the Northeast. I don't know. Like when I see these pictures, I don't think this is diverse enough}". These comments suggest that the models' disparagement of certain groups is not only a reflection of societal biases but also a perpetuation of harmful representations and erasure, which can have real-world consequences.
\section{Community-Driven Implications for Redesigning GAI Tools}

Drawing upon our participants' experiences and prior work in the field, we provide some insights around culturally-informed design principles for GAI tools such as T2Is. 

\subsection{Valuing the Evolution of Cultures}

A primary theme across participants' observations was how outputs embedded a sense of Indian culture that was either entirely inaccurate or, at the very least, reflective of an India from years gone by. As Arundhati stated \textit{``[the set of outputs] is very outdated, and not 2024,''} calling for a more updated representation of Indian culture and subcultures beyond stereotypical depictions from a \textit{``Western idea that's being portrayed for India, if so I think that should be changed" - Kamala}. Participants reasoned that while some of the exoticist and culturally misappropriated depictions of India were accurate to what things looked like at one point of time, such representations are no longer as synonymous to Indian-ness as these outputs might lead someone to believe. Indeed, they pointed to the fact that \textit{``a search engine results like Google Search [is] more representative of the Indian culture than what this is" - Anita}, implying that the excuse of a digital absence of comparably accurate representations or any other data to compare with for models to be moot. 

It is a central feature of culture that it changes and grows with time, with participants negotiating and renegotiating meanings of norms and objects \cite{hall1989cultural, hall1997representation_introduction}. While representations may originate from foundational depictions and create stereotypes which tend to endure \cite{hamilton}, these too must change as the passage of time and the evolution of a culture ascribes novel meaning to artefacts. This is especially true for colonized countries such as India as `Indian culture' today has an indelible imprint of its colonizers, and efforts to segregate which aspects of modern culture came from where might be fruitless \cite{foster1988culture}. Shedding depictions of a community heavily colored by colonial pasts requires a \textit{decolonial} perspective: an ``additive-inclusive approach, which continues to use existing knowledge, but in ways that recognises explicitly the value of new and alternative approaches, and that supports environments in which new ways of creating knowledge can genuinely flourish." \cite{mohamed2020decolonial} Our findings reveals a clear need for such an approach to overhauling representations of Indian culture within T2Is.   

This overhaul will not be easy: the volume of data on the Internet (typically used to train GAI models) that depicts India through a Western lens is quite significant, and despite the effects of globalization, there likely still are segments of the global population that ascribe to the stereotypes of Indian culture/subcultures documented in our findings. The overhaul would be most successful if the highest value is assigned to voices of those with the strongest epistemic experience in this regard: \textit{people who identify as Indian}. 

\subsection{Honoring Heterogeneity within Cultures}
However, as our participants demonstrated, Indian-ness is not a singular identity: there exist various subcultures with distinct features and traditions grounded in centuries of historical context. Our findings are full of instances where such diversity is homogenized, such as Satyajit's remarks that \textit{``[the output is] not capturing region-specific phenomena or region-specific features, [and] that's a drawback.''} Any efforts to depict Indian-ness within T2I model outputs must honor the heterogeneity and diversity of India. We align ourselves with \citet{qadri2023ai} in calling for stronger community-centered approaches towards designing T2I tools, and similarly advocate for the involvement of epistemic experts in the design process. Indian people are experts on their Indian-ness, and community-centered practices of data collection and annotation are an important step towards reducing harm within the outputs of T2I tools \cite{gadiraju2023wouldn, mack2024they}. 

However, asking individuals to be expert representatives of a culture/subculture has its own flaws, like the fact that this risks performing the same homogenization that this paper faults. Our participants recognized this dichotomy, and offered other ways of honoring heterogeneity within Indian culture. One such approach was the proposed fine-tuning of models for specific cultures/subcultures, instead of advertising them for widespread use. Participants noted that in trying to access a larger market, GAI tools risked causing harm to a wide group of users and losing a customer base. Taking inspiration from GAI tools such as Lesan, a machine translation tool and chatbot for Ethiopian languages which promises stronger results than tools from the Google Suite or ChatGPT \cite{lesan}, we advocate for stronger support and providing resources for designers of GAI tools for specific contexts, embedded within the culture they are trying to serve. Participants suggested that the models should acknowledge their limitations beyond broad statements such as `our results may be biased,' recognize the diversity of cultures and subcultures, provide disclaimers for inaccurate representations, and account for their impact. As Aravind effectively put it, `\textit{`It would have been good if it had a single line recognizing the fact that it's not composed of all the different cultures or subcultures that it's trying to describe.''}

\subsection{Accuracy and Cultural Sensitivity}
In analyzing behaviours of sociotechnical GAI models representing Indian subcultures, participants identified a unique facet related to model behavior diverging from expected actions, particularly concerning the given prompts. These deviations affect the model's accuracy within the lens of machine learning. The model creates erroneous responses based on the inputs, portraying them as factual wherein it was completely against what was asked for. An example of this phenomenon occurred when participants prompted the generation of images depicting Indian women separately in Western attire, modern wear, or non-traditional clothing. Despite these specific prompts, the model consistently generated images of women in traditional sarees, contrary to expectations. These examples highlight the need for GAI tools to accurately interpret and respond to prompts. Participants also emphasized the need for AI models to generate contextually appropriate responses that align with the prompt, demonstrate multilingual understanding by moving beyond weak transliterations and acknowledging diverse linguistic and cultural contexts, and avoid generating harmful content. 

In response to the concerns and harms outlined above, we provide a set of guidelines for creating culturally sensitive AI models, as shown in Figure \ref{fig:principles}. These guidelines, presented in detail in Appendix \ref{appendix-design}, provide a framework for developers and researchers to consider and adhere to, ensuring that their models prioritize cultural understanding, accuracy, and inclusivity. These considerations are rooted in our findings around Indian culture/subculture, but are adaptable for any underrepresented cultural group' towards more accurate representation of their contexts in GAI outputs. 

\begin{figure*}[h]
    \centering
    \includegraphics[scale=0.23]{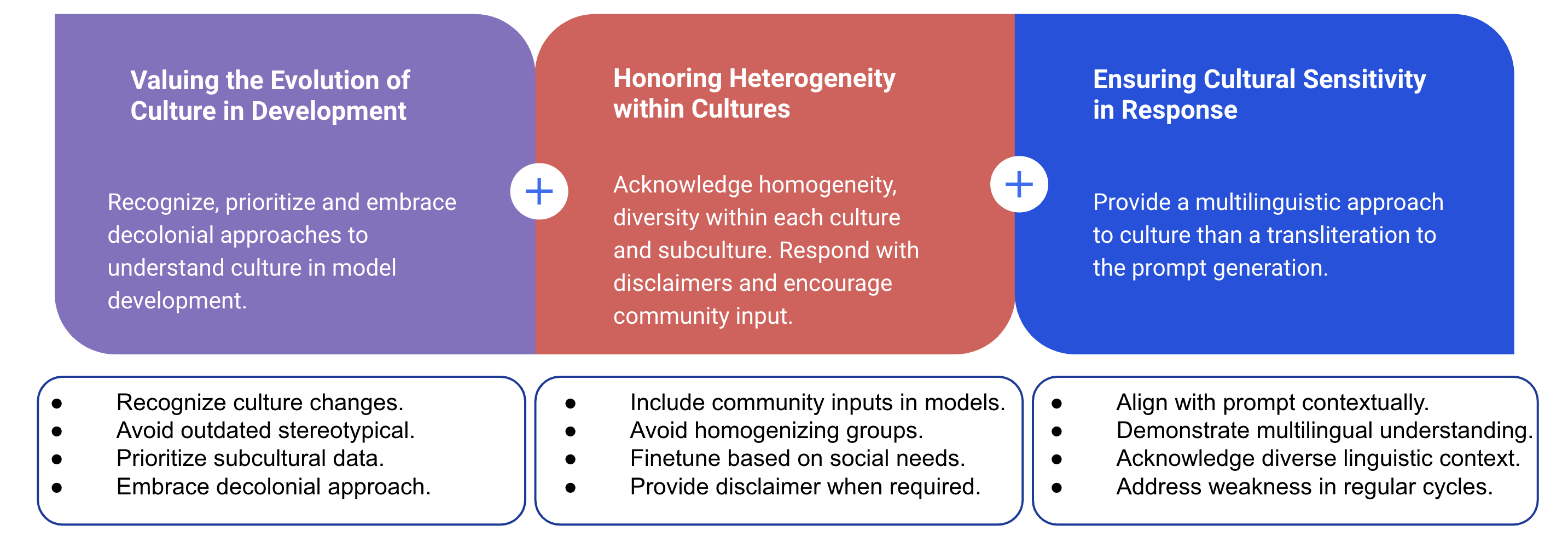}
    \caption{Design Recommendations based on community inputs for culturally sensitive development for T2I and GAI models.}
    \label{fig:principles}
\end{figure*}

\section{Discussion}

In this paper, we structure our narration around sociotechnical AI systems, exploring their impact on social and technical actors within specific ecosystems. These systems, exemplified by the widespread use of GAI tools in domains domains that impact society and people's life outcomes and opportunitie, such as education and policy-making, often operate as opaque `black boxes' \cite{o2017weapons}. This lack of transparency means that many users need to be made aware of the underlying mechanisms and decision-making processes of these models, leading to instances of discrimination in social contexts. Recognizing this, our research was motivated by the need for a community-based study focused on  GAI and sociotechnical systems for public consumption.

As the focus group progressed, a clear consensus emerged among participants: the current state of text and image generation models is inadequate and often harmful. The models' outputs were frequently deemed \textit{toxic, hateful, stereotypical,} or \textit{outright wrong,} reflecting a post-colonial and Western viewpoint that failed to resonate with the community. This disconnect is not surprising, given the historical and ongoing impact of colonialism on people's identities, as highlighted by \cite{das2024colonial}. There appears to be \textit{'branded diversity through tokenism'} \cite{stevens2022black} when it comes to representation of the Indian cultural context. The perpetuation of biases through sociotechnical systems is a critical concern. Our study reveals how these systems can amplify harmful stereotypes, contributing to a cycle of cultural misappropriation and erasure. The participants' desire for better results is not just a call for technical improvement but also a demand for more inclusive and culturally sensitive AI systems.

Our work addresses a significant gap in the field by centering a Global South context, taking away the focus from Westernized sites of study and lenses of examining harm, to explore cultural harm and its nuances within various subcultures \cite{sambasivan2021re}. We use the case study of India, an especially relevant context given a rising concern around AI-mediated harms there, as GAI-generated deepfakes are being used to garner votes in elections from viewers who are largely unaware of such videos being synthetically generated \cite{nyt_raj}. However, our novel harms of exoticism and cultural misappropriation can be further expanded to the contexts of Global South and other contexts not centered in GAI design. This possibly applies the strongest to non-Western cultures, as evidenced by GAI depictions of African houses, even those of `wealthy' African people, as rural huts \cite{bianchi2023easily}. This depiction, and the fact that the addition of the word `wealthy' made no difference in receiving huts within the outputs, shows how the representational harms of exoticism and cultural appropriation might apply to other cultures. Depictions of such cultures, might be similarly marred by a Westernized lens likely steeped in colonialism and imperialism. It is essential to recognize and address the ways such impacts of colonialism within the design of globally-prevalent GAI tools has shaped and thus begin to develop GAI systems that prioritize the perspectives of diverse communities.  

Towards this end, our framework (Figure \ref{fig:principles}) proposes \textit{key considerations} for developing culturally sensitive GAI tools. In spite of the fact that such models will likely be ubiquitous and embedded within suites of AIaaS, we do not advocate for theories of universal design that seek to produce \textit{one-size-fits-all} solutions. Instead, rooted in participant observations, we advocate for fine-grained tools for specific contexts, and focus on designing for non-Western cultures.   

\section{Limitations, Future Work and Conclusion}

One limitation of our work is that even though we present evidence on the erasure of certain Indian subcultures in generative AI results, we do not prominently feature voices in this study of people identifying with those subcultures, in that we only have one participant identifying from Northeastern India and no Muslim participants. This is an unfortunate consequence of recruitment practices and response to our offers, and a deliberate decision to not specifically seek out users of any particular identity within our target population so as to avoid introducing researcher bias into our data and to not ask marginalized users to perform their marginalization for our extractive research purposes. We also present evidence of participants labeling human faces depicted within Stable Diffusion outputs as appearing to be from specific parts of India, such as the North. While we respect those inferences since they are rooted in participants' positionalities and lived experiences within their cultural context, we note the possibility of readers disagreeing with classifications based on their own perceptions of what it means for someone to `look' a particular type of Indian. 

Furthermore, this study is conducted using English prompts asking for representations of Indian culture. Although English is widely spoken in India and a lot of GAI usage is done so in English, future iterations of this study can experiment with prompts in Indian languages. 

Our work highlights the pressing issue of cultural harms within GAI, particularly in non-Western contexts such as India. Through community-centric research that engages with diverse Indian communities, we identified novel forms of representational harm, exoticism and cultural misappropriation on top of the existing harms \cite{dev2020measuring}. We propose design principles prioritizing cultural diversity, and respect to mitigate the perpetuation of harmful misrepresentations. By building upon existing advocacy for ethical GAI design in non-Western contexts and drawing parallels with similar challenges faced by other cultures, our work contributes to showcasing the importance of cultural awareness and sensitivity in the development of AI technologies.

\section*{Positionality Statement}

We situate ourselves within this research by acknowledging our own positionalities, which shape our perspectives and biases \cite{haraway1988situated}. The three co-authors of this paper researchers identify as being born and brought up in India, having spent their formative years in India, and the fourth author identifies as a second-generation Indian. The fifth author identifies as an American of Bulgarian-Turkish origin, having spent half of their life in the United States. Although all authors currently reside in the United States, our collective goal and research efforts focus on investigating biases related to identity in global, multilingual, and culturally inclusive language technology. Our focus on critical HCI, marginalized communities, and ethnolinguistic groups drives our exploration and collaboration. Through all our prior experience, we bring both understandings of non-western contexts and experiences while also recognizing our privileged positions within academia in the Global North.

\section*{Ethical Consideration}
Our research into the representations of non-Western cultures within Generative Artificial Intelligence outputs, specifically Text-to-Image models, focusing on Indian cultural contexts, is guided by the commitment to ethical considerations and the impact they can have on the community of focus. We recognize the potential impacts of GAI tools on cultural narratives and identities and conduct our work in a manner that respects the dignity, diversity, and complexity of cultural representations. We prioritize cultural respect, avoiding harm through biased or misrepresentative outputs, and inclusivity by engaging diverse voices and perspectives. Our methodologies are transparent and accountable, and we empower communities to participate in discussions about AI ethics and fairness. Moreover, we adopt a community-centric research approach, prioritizing participant experience and enabling communities to lead and drive the conversation, structuring a more ethnographic approach to our results. We have made our study's research materials publicly available\footnote{https://github.com/PranavNV/Harms\_In\_GAI} to promote transparency and facilitate further research. Our intention is to encourage additional investigation and build upon our findings, ultimately contributing to more inclusive and diverse cultural representation in the current generative AI landscape. By upholding these principles, we aim to conduct research that is respectful, inclusive, and ethical, contributing to a more responsible development of GAI tools. 

\section*{Adverse Impact Statement} 

By focusing on the representation of non-Western cultures, specifically Indian culture, the project exposes critical issues related to biased outputs and cultural misappropriation within AI-generated content. The identified harms, including exoticism and cultural misappropriation, highlight significant challenges in ensuring fair and accurate representations of diverse cultural contexts. The project's findings reveal the inherent biases embedded within GAI systems, which can perpetuate stereotypes, misrepresentations, and cultural erasure. These biases not only undermine the richness and complexity of non-Western cultures but also contribute to broader societal issues such as marginalization and stereotyping. Furthermore, the persistence of exoticist portrayals and cultural misappropriation underscores the systemic nature of these biases within AI technologies. Such adverse impacts can have far-reaching consequences, influencing how individuals perceive and interact with different cultures, perpetuating harmful stereotypes, and reinforcing power imbalances. Addressing these adverse impacts requires a concerted effort from researchers, developers, and policymakers to prioritize ethical considerations, mitigate biases, and ensure responsible AI development and deployment. Failure to address these issues may result in continued harm, perpetuation of stereotypes, and exacerbation of cultural inequalities, ultimately hindering the potential benefits of GAI tools for diverse global communities.

\section*{Acknowledgements}
This work was supported by the U.S. National Institute of Standards and Technology (NIST) Grant 60NANB23D194. Any opinions, findings, and conclusions or recommendations expressed in this material are those of the authors and do not necessarily reflect those of NIST.

We are also appreciative of our reviewers for their thoughtful feedback, which undoubtedly made this paper richer than it otherwise would have been. 

\bibliography{aaai24.bib}

\begin{thebibliography}{81}
\providecommand{\natexlab}[1]{#1}

\bibitem[{Arora and Sylvia(2023)}]{arora2023just}
Arora, A.; and Sylvia, N.~P. 2023.
\newblock “Just Like Everyone Else:” queer representation in postmillennial bollywood.
\newblock \emph{Feminist Media Studies}, 1--15.

\bibitem[{Babar(2024)}]{Babar24}
Babar, K. 2024.
\newblock Mumbai breaks into world’s top 10 luxury residential markets, now ranks 8th.

\bibitem[{Barocas et~al.(2017)Barocas, Crawford, Shapiro, and Wallach}]{barocas2017problem}
Barocas, S.; Crawford, K.; Shapiro, A.; and Wallach, H. 2017.
\newblock The problem with bias: Allocative versus representational harms in machine learning.
\newblock In \emph{9th Annual conference of the Special Interest Group for Computing, Information and Society}.

\bibitem[{Bender et~al.(2021)Bender, Gebru, McMillan-Major, and Shmitchell}]{bender2021dangers}
Bender, E.~M.; Gebru, T.; McMillan-Major, A.; and Shmitchell, S. 2021.
\newblock On the dangers of stochastic parrots: Can language models be too big?
\newblock In \emph{Proceedings of the 2021 ACM conference on fairness, accountability, and transparency}, 610--623.

\bibitem[{Benedict(2024)}]{benedict2024deconstructing}
Benedict, S.~M. 2024.
\newblock Deconstructing Notions of Women's Attire: Challenging Stereotypes, Unveiling Hypocrisy, and Decoding Dichotomies.
\newblock \emph{Unveiling Hypocrisy, and Decoding Dichotomies (March 1, 2024)}.

\bibitem[{Bhat(2019)}]{bhat2019muslim}
Bhat, S.~H. 2019.
\newblock Muslim Characters in Bollywood Cinema: Representation and Reality.
\newblock \emph{IOSR Journal of Humanities and Social Science}, 24(12).

\bibitem[{Bianchi et~al.(2023)Bianchi, Kalluri, Durmus, Ladhak, Cheng, Nozza, Hashimoto, Jurafsky, Zou, and Caliskan}]{bianchi2023easily}
Bianchi, F.; Kalluri, P.; Durmus, E.; Ladhak, F.; Cheng, M.; Nozza, D.; Hashimoto, T.; Jurafsky, D.; Zou, J.; and Caliskan, A. 2023.
\newblock Easily accessible text-to-image generation amplifies demographic stereotypes at large scale.
\newblock In \emph{Proceedings of the 2023 ACM Conference on Fairness, Accountability, and Transparency}, 1493--1504.

\bibitem[{Birhane(2021)}]{birhane2021algorithmic}
Birhane, A. 2021.
\newblock Algorithmic injustice: a relational ethics approach.
\newblock \emph{Patterns}, 2(2).

\bibitem[{Blodgett(2021)}]{blodgett2021sociolinguistically}
Blodgett, S.~L. 2021.
\newblock Sociolinguistically driven approaches for just natural language processing.

\bibitem[{Bolukbasi et~al.(2016)Bolukbasi, Chang, Zou, Saligrama, and Kalai}]{bolukbasi2016man}
Bolukbasi, T.; Chang, K.-W.; Zou, J.~Y.; Saligrama, V.; and Kalai, A.~T. 2016.
\newblock Man is to computer programmer as woman is to homemaker? debiasing word embeddings.
\newblock \emph{Advances in neural information processing systems}, 29.

\bibitem[{Brynjolfsson, Li, and Raymond(2023)}]{brynjolfsson2023generative}
Brynjolfsson, E.; Li, D.; and Raymond, L.~R. 2023.
\newblock Generative AI at work.
\newblock Technical report, National Bureau of Economic Research.

\bibitem[{Caliskan, Bryson, and Narayanan(2017)}]{caliskan2017semantics}
Caliskan, A.; Bryson, J.~J.; and Narayanan, A. 2017.
\newblock Semantics derived automatically from language corpora contain human-like biases.
\newblock \emph{Science}, 356(6334): 183--186.

\bibitem[{Chakkarath(2010)}]{chakkarath2010stereotypes}
Chakkarath, P. 2010.
\newblock Stereotypes in social psychology: The “West-East “differentiation as a reflection of Western traditions of thought.
\newblock \emph{Psychological Studies}, 55: 18--25.

\bibitem[{Chandramouli(2011)}]{indiacensus}
Chandramouli, C. 2011.
\newblock 2011 Census Report.
\newblock Technical report, Government of India.

\bibitem[{Charmaz(2006)}]{charmaz2006constructing}
Charmaz, K. 2006.
\newblock \emph{Constructing grounded theory: A practical guide through qualitative analysis}.
\newblock sage.

\bibitem[{Charmaz(2017)}]{charmaz2017constructivist}
Charmaz, K. 2017.
\newblock Constructivist grounded theory.
\newblock \emph{The Journal of Positive Psychology}, 12(3): 299--300.

\bibitem[{Chaudhuri(2009)}]{chaudhuri2009snake}
Chaudhuri, S. 2009.
\newblock Snake charmers and child brides: Deepa Mehta's Water,‘exotic’representation, and the cross-cultural spectatorship of South Asian migrant cinema.
\newblock \emph{South Asian Popular Culture}, 7(1): 7--20.

\bibitem[{Cooper and Foster(1971)}]{cooper1971sociotechnical}
Cooper, R.; and Foster, M. 1971.
\newblock Sociotechnical systems.
\newblock \emph{American Psychologist}, 26(5): 467.

\bibitem[{Das et~al.(2024)Das, Guha, Brubaker, and Semaan}]{das2024colonial}
Das, D.; Guha, S.; Brubaker, J.; and Semaan, B. 2024.
\newblock The" Colonial Impulse" of Natural Language Processing: An Audit of Bengali Sentiment Analysis Tools and Their Identity-based Biases.
\newblock \emph{arXiv preprint arXiv:2401.10535}.

\bibitem[{Deck(2023)}]{lesan}
Deck, A. 2023.
\newblock The AI startup outperforming Google Translate in Ethiopian languages.
\newblock \emph{Rest of World}.

\bibitem[{Deepshikha, Yammiyavar, and Nath(2018)}]{deepshikha2018textiles}
Deepshikha; Yammiyavar, P.; and Nath, N. 2018.
\newblock Textiles as communicating links for cultural traditions.
\newblock In \emph{Proceedings of the 7th International Conference on Kansei Engineering and Emotion Research 2018: KEER 2018, 19-22 March 2018, Kuching, Sarawak, Malaysia}, 168--177. Springer.

\bibitem[{Dev et~al.(2020)Dev, Li, Phillips, and Srikumar}]{dev2020measuring}
Dev, S.; Li, T.; Phillips, J.~M.; and Srikumar, V. 2020.
\newblock On measuring and mitigating biased inferences of word embeddings.
\newblock In \emph{Proceedings of the AAAI Conference on Artificial Intelligence}, volume~34, 7659--7666.

\bibitem[{Dev et~al.(2022)Dev, Sheng, Zhao, Amstutz, Sun, Hou, Sanseverino, Kim, Nishi, Peng et~al.}]{dev2022measures}
Dev, S.; Sheng, E.; Zhao, J.; Amstutz, A.; Sun, J.; Hou, Y.; Sanseverino, M.; Kim, J.; Nishi, A.; Peng, N.; et~al. 2022.
\newblock On Measures of Biases and Harms in NLP.
\newblock In \emph{Findings of the Association for Computational Linguistics: AACL-IJCNLP 2022}, 246--267.

\bibitem[{D'ignazio and Klein(2020)}]{d2020data}
D'ignazio, C.; and Klein, L.~F. 2020.
\newblock \emph{Data feminism}.
\newblock MIT press.

\bibitem[{Dowerah and Nath(2017)}]{dowerah2017cinematic}
Dowerah, S.; and Nath, D.~P. 2017.
\newblock Cinematic regimes of otherness: India and its Northeast.
\newblock \emph{Media Asia}, 44(2): 121--133.

\bibitem[{Field et~al.(2021)Field, Blodgett, Waseem, and Tsvetkov}]{field2021survey}
Field, A.; Blodgett, S.~L.; Waseem, Z.; and Tsvetkov, Y. 2021.
\newblock A survey of race, racism, and anti-racism in NLP.
\newblock \emph{arXiv preprint arXiv:2106.11410}.

\bibitem[{Foster(1988)}]{foster1988culture}
Foster, J.~W. 1988.
\newblock Culture and colonisation: view from the North.
\newblock \emph{The Irish Review (1986-)}, (5): 17--26.

\bibitem[{Gadiraju et~al.(2023)Gadiraju, Kane, Dev, Taylor, Wang, Denton, and Brewer}]{gadiraju2023wouldn}
Gadiraju, V.; Kane, S.; Dev, S.; Taylor, A.; Wang, D.; Denton, E.; and Brewer, R. 2023.
\newblock " I wouldn’t say offensive but...": Disability-Centered Perspectives on Large Language Models.
\newblock In \emph{Proceedings of the 2023 ACM Conference on Fairness, Accountability, and Transparency}, 205--216.

\bibitem[{Gautam, Venkit, and Ghosh(2024)}]{gautam2024melting}
Gautam, S.; Venkit, P.~N.; and Ghosh, S. 2024.
\newblock From Melting Pots to Misrepresentations: Exploring Harms in Generative AI.
\newblock \emph{arXiv preprint arXiv:2403.10776}.

\bibitem[{Glaser(1992)}]{glaser1992basics}
Glaser, B. 1992.
\newblock Basics of grounded theory analysis: Emergence vs forcing.

\bibitem[{Glaser and Strauss(1967)}]{glaser1967discovery}
Glaser, B.; and Strauss, A. 1967.
\newblock \emph{Discovery of grounded theory: Strategies for qualitative research}.
\newblock Routledge.

\bibitem[{Gupta et~al.(2023{\natexlab{a}})Gupta, Venkit, Lauren{\c{c}}on, Wilson, and Passonneau}]{gupta2023calm}
Gupta, V.; Venkit, P.~N.; Lauren{\c{c}}on, H.; Wilson, S.; and Passonneau, R.~J. 2023{\natexlab{a}}.
\newblock CALM: A Multi-task Benchmark for Comprehensive Assessment of Language Model Bias.
\newblock \emph{arXiv preprint arXiv:2308.12539}.

\bibitem[{Gupta et~al.(2023{\natexlab{b}})Gupta, Venkit, Wilson, and Passonneau}]{gupta2023sociodemographic}
Gupta, V.; Venkit, P.~N.; Wilson, S.; and Passonneau, R.~J. 2023{\natexlab{b}}.
\newblock Sociodemographic Bias in Language Models: A Survey and Forward Path.
\newblock \emph{arXiv e-prints}, arXiv--2306.

\bibitem[{Halcomb et~al.(2007)Halcomb, Gholizadeh, DiGiacomo, Phillips, and Davidson}]{halcomb2007literature}
Halcomb, E.~J.; Gholizadeh, L.; DiGiacomo, M.; Phillips, J.; and Davidson, P.~M. 2007.
\newblock Literature review: considerations in undertaking focus group research with culturally and linguistically diverse groups.
\newblock \emph{Journal of clinical nursing}, 16(6): 1000--1011.

\bibitem[{Hall(1989)}]{hall1989cultural}
Hall, S. 1989.
\newblock Cultural identity and cinematic representation.
\newblock \emph{Framework: The Journal of Cinema and Media}, (36): 68--81.

\bibitem[{Hall(1997{\natexlab{a}})}]{hall1997representation}
Hall, S. 1997{\natexlab{a}}.
\newblock \emph{Representation: Cultural representations and signifying practices}, volume~2.
\newblock Sage.

\bibitem[{Hall(1997{\natexlab{b}})}]{hall1997representation_introduction}
Hall, S. 1997{\natexlab{b}}.
\newblock \emph{Representation: cultural representations and signifying practices}, chapter~i, 1--12.
\newblock Milton Keynes, UK: The Open University.

\bibitem[{Hamilton(1997)}]{hamilton}
Hamilton, P. 1997.
\newblock Representing the social: France and frenchness in post-war humanist photography.
\newblock In Hall, S., ed., \emph{Representation: Cultural representations and signifying practices}.

\bibitem[{Haraway(1988)}]{haraway1988situated}
Haraway, D. 1988.
\newblock ‘Situated Knowledges: The Science Question in Feminism and the Privilege of Partial Perspective'.
\newblock In \emph{Space, Gender, Knowledge: Feminist Readings}, 53--72. Routledge.

\bibitem[{Hasan(2011)}]{hasan2011talking}
Hasan, D. 2011.
\newblock Talking Back to ‘Bollywood’: Hindi Commercial Cinema in North-East India.
\newblock \emph{South Asian Media Cultures. Audiences, Representations, Contexts, London}.

\bibitem[{Hastings(2024)}]{hastings2024preventing}
Hastings, J. 2024.
\newblock Preventing harm from non-conscious bias in medical generative AI.
\newblock \emph{The Lancet Digital Health}, 6(1): e2--e3.

\bibitem[{Huer and Saenz(2003)}]{huer2003challenges}
Huer, M.~B.; and Saenz, T.~I. 2003.
\newblock Challenges and strategies for conducting survey and focus group research with culturally diverse groups.

\bibitem[{Hughes and DuMont(1993)}]{hughes1993using}
Hughes, D.; and DuMont, K. 1993.
\newblock Using focus groups to facilitate culturally anchored research.
\newblock \emph{American journal of community psychology}, 21(6): 775--806.

\bibitem[{Islam(2007)}]{islam2007imagining}
Islam, M. 2007.
\newblock Imagining Indian Muslims: Looking through the lens of Bollywood cinema.
\newblock \emph{Indian Journal of Human Development}, 1(2): 403--422.

\bibitem[{Khadilkar, KhudaBukhsh, and Mitchell(2022)}]{khadilkar2022gender}
Khadilkar, K.; KhudaBukhsh, A.~R.; and Mitchell, T.~M. 2022.
\newblock Gender bias, social bias, and representation in Bollywood and Hollywood.
\newblock \emph{Patterns}, 3(2): 100409.

\bibitem[{Kotliar(2020)}]{kotliar2020data}
Kotliar, D.~M. 2020.
\newblock Data orientalism: on the algorithmic construction of the non-Western other.
\newblock \emph{Theory and Society}, 49(5): 919--939.

\bibitem[{Kumar(2013)}]{kumar2013constructing}
Kumar, S. 2013.
\newblock Constructing the Nation’s Enemy: Hindutva, popular culture and the Muslim ‘other’in Bollywood cinema.
\newblock \emph{Third World Quarterly}, 34(3): 458--469.

\bibitem[{Lewicki et~al.(2023)Lewicki, Lee, Cobbe, and Singh}]{lewicki2023out}
Lewicki, K.; Lee, M. S.~A.; Cobbe, J.; and Singh, J. 2023.
\newblock Out of Context: Investigating the Bias and Fairness Concerns of “Artificial Intelligence as a Service”.
\newblock In \emph{Proceedings of the 2023 CHI Conference on Human Factors in Computing Systems}, 1--17.

\bibitem[{MacDougall and Baum(1997)}]{macdougall1997devil}
MacDougall, C.; and Baum, F. 1997.
\newblock The devil's advocate: A strategy to avoid groupthink and stimulate discussion in focus groups.
\newblock \emph{Qualitative health research}, 7(4): 532--541.

\bibitem[{Mack et~al.(2024)Mack, Qadri, Denton, Kane, and Bennett}]{mack2024they}
Mack, A.; Qadri, R.; Denton, R.; Kane, S.~K.; and Bennett, C.~L. 2024.
\newblock ``{T}hey only care to show us the wheelchair”: Disability Representation in text-to-image {AI} models.
\newblock In \emph{Proceedings of the 2024 CHI Conference on Human Factors in Computing Systems}.

\bibitem[{Mamdani(2002)}]{mamdani2002good}
Mamdani, M. 2002.
\newblock Good Muslim, bad Muslim: A political perspective on culture and terrorism.
\newblock \emph{American anthropologist}, 104(3): 766--775.

\bibitem[{Matusitz and Payano(2011)}]{matusitz2011bollywood}
Matusitz, J.; and Payano, P. 2011.
\newblock The Bollywood in Indian and American perceptions: A comparative analysis.
\newblock \emph{India Quarterly}, 67(1): 65--77.

\bibitem[{Matusitz and Payano(2012)}]{matusitz2012globalisation}
Matusitz, J.; and Payano, P. 2012.
\newblock Globalisation of popular culture: From Hollywood to Bollywood.
\newblock \emph{South Asia Research}, 32(2): 123--138.

\bibitem[{Mehta and Pandharipande(2011)}]{mehta2011bollywood}
Mehta, R.~B.; and Pandharipande, R.~V. 2011.
\newblock \emph{Bollywood and globalization: Indian popular cinema, nation, and diaspora}.
\newblock Anthem Press.

\bibitem[{Mohamed, Png, and Isaac(2020)}]{mohamed2020decolonial}
Mohamed, S.; Png, M.-T.; and Isaac, W. 2020.
\newblock Decolonial AI: Decolonial theory as sociotechnical foresight in artificial intelligence.
\newblock \emph{Philosophy \& Technology}, 33: 659--684.

\bibitem[{O'neil(2017)}]{o2017weapons}
O'neil, C. 2017.
\newblock \emph{Weapons of math destruction: How big data increases inequality and threatens democracy}.
\newblock Crown.

\bibitem[{Panda and Gupta(2004)}]{panda2004mapping}
Panda, A.; and Gupta, R.~K. 2004.
\newblock Mapping cultural diversity within India: A meta-analysis of some recent studies.
\newblock \emph{Global Business Review}, 5(1): 27--49.

\bibitem[{Pattnaik and Samantaray(2017)}]{pattnaik2017comparative}
Pattnaik, S.; and Samantaray, S.~P. 2017.
\newblock Comparative study on classical dances: Odissi \& Bharatanatyam.
\newblock \emph{World Wide J Multidiscip Res Dev}, 3: 147--150.

\bibitem[{Peters(2021)}]{peters2021colorism}
Peters, R. 2021.
\newblock Colorism, castism, and gentrification in Bollywood.
\newblock \emph{The Jugaad Project}, 24.

\bibitem[{Prabhakaran, Qadri, and Hutchinson(2022)}]{prabhakaran2022cultural}
Prabhakaran, V.; Qadri, R.; and Hutchinson, B. 2022.
\newblock Cultural incongruencies in artificial intelligence.
\newblock \emph{arXiv preprint arXiv:2211.13069}.

\bibitem[{Qadri et~al.(2023)Qadri, Shelby, Bennett, and Denton}]{qadri2023ai}
Qadri, R.; Shelby, R.; Bennett, C.~L.; and Denton, E. 2023.
\newblock AI’s Regimes of Representation: A Community-centered Study of Text-to-Image Models in South Asia.
\newblock In \emph{Proceedings of the 2023 ACM Conference on Fairness, Accountability, and Transparency}, 506--517.

\bibitem[{Raj(2024)}]{nyt_raj}
Raj, S. 2024.
\newblock How A.I. Tools Could Change India’s Elections.
\newblock \emph{The New York Times}.

\bibitem[{Rane et~al.(2023)Rane, Tawde, Choudhary, and Rane}]{rane2023contribution}
Rane, N.~L.; Tawde, A.; Choudhary, S.~P.; and Rane, J. 2023.
\newblock Contribution and performance of ChatGPT and other Large Language Models (LLM) for scientific and research advancements: a double-edged sword.
\newblock \emph{International Research Journal of Modernization in Engineering Technology and Science}, 5(10): 875--899.

\bibitem[{Ray(1958)}]{ray1988}
Ray, S. 1958.
\newblock Problems of a Bengali Film Maker.
\newblock \emph{International Film Annual}.

\bibitem[{Rodriguez et~al.(2011)Rodriguez, Schwartz, Lahman, and Geist}]{rodriguez2011culturally}
Rodriguez, K.~L.; Schwartz, J.~L.; Lahman, M.~K.; and Geist, M.~R. 2011.
\newblock Culturally responsive focus groups: Reframing the research experience to focus on participants.
\newblock \emph{International Journal of Qualitative Methods}, 10(4): 400--417.

\bibitem[{Said(1977)}]{said1977orientalism}
Said, E.~W. 1977.
\newblock Orientalism.
\newblock \emph{The Georgia Review}, 31(1): 162--206.

\bibitem[{Sambasivan et~al.(2021{\natexlab{a}})Sambasivan, Arnesen, Hutchinson, Doshi, and Prabhakaran}]{sambasivan2021re}
Sambasivan, N.; Arnesen, E.; Hutchinson, B.; Doshi, T.; and Prabhakaran, V. 2021{\natexlab{a}}.
\newblock Re-imagining algorithmic fairness in india and beyond.
\newblock In \emph{Proceedings of the 2021 ACM conference on fairness, accountability, and transparency}, 315--328.

\bibitem[{Sambasivan et~al.(2021{\natexlab{b}})Sambasivan, Kapania, Highfill, Akrong, Paritosh, and Aroyo}]{sambasivan2021everyone}
Sambasivan, N.; Kapania, S.; Highfill, H.; Akrong, D.; Paritosh, P.; and Aroyo, L.~M. 2021{\natexlab{b}}.
\newblock “Everyone wants to do the model work, not the data work”: Data Cascades in High-Stakes AI.
\newblock In \emph{proceedings of the 2021 CHI Conference on Human Factors in Computing Systems}, 1--15.

\bibitem[{Sap et~al.(2019)Sap, Card, Gabriel, Choi, and Smith}]{sap2019risk}
Sap, M.; Card, D.; Gabriel, S.; Choi, Y.; and Smith, N.~A. 2019.
\newblock The risk of racial bias in hate speech detection.
\newblock In \emph{Proceedings of the 57th annual meeting of the association for computational linguistics}, 1668--1678.

\bibitem[{Sen(1997)}]{sen1997indian}
Sen, A. 1997.
\newblock Indian traditions and the Western imagination.
\newblock \emph{Daedalus}, 126(2): 1--26.

\bibitem[{Septiandri et~al.(2023)Septiandri, Constantinides, Tahaei, and Quercia}]{septiandri2023weird}
Septiandri, A.~A.; Constantinides, M.; Tahaei, M.; and Quercia, D. 2023.
\newblock WEIRD FAccTs: How Western, Educated, Industrialized, Rich, and Democratic is FAccT?
\newblock In \emph{Proceedings of the 2023 ACM Conference on Fairness, Accountability, and Transparency}, 160--171.

\bibitem[{Shelby et~al.(2022)Shelby, Rismani, Henne, Moon, Rostamzadeh, Nicholas, Yilla, Gallegos, Smart, Garcia et~al.}]{shelby2022sociotechnical}
Shelby, R.; Rismani, S.; Henne, K.; Moon, A.; Rostamzadeh, N.; Nicholas, P.; Yilla, N.; Gallegos, J.; Smart, A.; Garcia, E.; et~al. 2022.
\newblock Sociotechnical harms: Scoping a taxonomy for harm reduction. arXiv.

\bibitem[{Stevens(2022)}]{stevens2022black}
Stevens, W.~E. 2022.
\newblock \emph{Black Influencers: Interrogating the Racialization and Commodification of Digital Labor}.
\newblock Temple University.

\bibitem[{Trivedi(2008)}]{trivedi2008bollywood}
Trivedi, H. 2008.
\newblock From Bollywood to Hollywood: the globalization of Hindi cinema.
\newblock \emph{The Postcolonial and the Global}, 200--210.

\bibitem[{Venkit et~al.(2023{\natexlab{a}})Venkit, Srinath, Gautam, Venkatraman, Gupta, Passonneau, and Wilson}]{venkit2023sentiment}
Venkit, P.; Srinath, M.; Gautam, S.; Venkatraman, S.; Gupta, V.; Passonneau, R.~J.; and Wilson, S. 2023{\natexlab{a}}.
\newblock The Sentiment Problem: A Critical Survey towards Deconstructing Sentiment Analysis.
\newblock In \emph{Proceedings of the 2023 Conference on Empirical Methods in Natural Language Processing}, 13743--13763.

\bibitem[{Venkit(2023)}]{narayanan2023towards}
Venkit, P.~N. 2023.
\newblock Towards a Holistic Approach: Understanding Sociodemographic Biases in NLP Models using an Interdisciplinary Lens.
\newblock In \emph{Proceedings of the 2023 AAAI/ACM Conference on AI, Ethics, and Society}, 1004--1005.

\bibitem[{Venkit et~al.(2024)Venkit, Chakravorti, Gupta, Biggs, Srinath, Goswami, Rajtmajer, and Wilson}]{venkit2024confidently}
Venkit, P.~N.; Chakravorti, T.; Gupta, V.; Biggs, H.; Srinath, M.; Goswami, K.; Rajtmajer, S.; and Wilson, S. 2024.
\newblock " Confidently Nonsensical?'': A Critical Survey on the Perspectives and Challenges of'Hallucinations' in NLP.
\newblock \emph{arXiv preprint arXiv:2404.07461}.

\bibitem[{Venkit et~al.(2023{\natexlab{b}})Venkit, Gautam, Panchanadikar, Huang, and Wilson}]{venkit2023nationality}
Venkit, P.~N.; Gautam, S.; Panchanadikar, R.; Huang, T.-H.; and Wilson, S. 2023{\natexlab{b}}.
\newblock Nationality Bias in Text Generation.
\newblock In \emph{Proceedings of the 17th Conference of the European Chapter of the Association for Computational Linguistics}, 116--122.

\bibitem[{Venkit et~al.(2023{\natexlab{c}})Venkit, Gautam, Panchanadikar, Huang, and Wilson}]{narayanan2023unmasking}
Venkit, P.~N.; Gautam, S.; Panchanadikar, R.; Huang, T.-H.; and Wilson, S. 2023{\natexlab{c}}.
\newblock Unmasking nationality bias: A study of human perception of nationalities in ai-generated articles.
\newblock In \emph{Proceedings of the 2023 AAAI/ACM Conference on AI, Ethics, and Society}, 554--565.

\bibitem[{Venkit, Srinath, and Wilson(2023)}]{venkit2023automated}
Venkit, P.~N.; Srinath, M.; and Wilson, S. 2023.
\newblock Automated Ableism: An Exploration of Explicit Disability Biases in Sentiment and Toxicity Analysis Models.
\newblock In \emph{Proceedings of the 3rd Workshop on Trustworthy Natural Language Processing (TrustNLP 2023)}, 26--34.

\bibitem[{Zack et~al.(2023)Zack, Lehman, Suzgun, Rodriguez, Celi, Gichoya, Jurafsky, Szolovits, Bates, Abdulnour et~al.}]{zack2023coding}
Zack, T.; Lehman, E.; Suzgun, M.; Rodriguez, J.~A.; Celi, L.~A.; Gichoya, J.; Jurafsky, D.; Szolovits, P.; Bates, D.~W.; Abdulnour, R.-E.~E.; et~al. 2023.
\newblock Coding Inequity: Assessing GPT-4's Potential for Perpetuating Racial and Gender Biases in Healthcare.
\newblock \emph{medRxiv}, 2023--07.

\end{thebibliography}
\newpage
\appendix
\section{Appendix}
In this section, we provide complementary resources and experimental information that were omitted from the primary narration of the paper. This includes additional visualizations and a detailed description of the publicly accessible repository containing our codebook and the comprehensive collection of images generated during the focus groups. This supplementary material is intended to provide further context and transparency, supporting the reproducibility and replicability of our research.

\subsection{Community-Based Design Recommendations} \label{appendix-design}
Drawing from our study outcomes and the insights shared by our participants during our various focus group discussions, we have meticulously curated a set of thematic design principles that encapsulate key recommendations for enhancing model interactions in future developments. These guidelines are aimed at shaping the development of safer and culturally sensitive GAI systems, fostering mindfulness and inclusivity in their design.\\

\textbf{I. Valuing the Evolution of Cultures:}
\begin{enumerate}
\item Recognize that cultures change and grow over time.

\item Avoid outdated and stereotypical representations.
    
\item Prioritize accurate and updated representations of a community's culture and subcultures.
    
\item Embrace a decolonial approach to overhauling representations.
\end{enumerate}
\textbf{II. Honoring Heterogeneity within Cultures}
\begin{enumerate}
\item Acknowledge the diversity within a culture.

\item Avoid homogenizing a group, in this case,  Indian-ness.

\item Involve epistemic experts in the design process to provide sound inputs on the group in focus.

\item Consider fine-tuning models for specific cultures/subcultures based on their needs.

\item Provide disclaimers for inaccurate representations wherever possible.
\end{enumerate}

\textbf{III. Ensuring Accuracy and Cultural Sensitivity \\in Responses}
\begin{enumerate}
\item Generate contextually appropriate responses that align with the prompt.

\item Demonstrate multilingual understanding by moving beyond weak transliterations.

\item Acknowledge diverse linguistic context within a cultural group.

\item Address weaknesses and avoid generating harmful or wrongful content.
\end{enumerate}

\subsection{Additional Images}\label{appendix-images}
\subsection{Participant Details}

\begin{figure}[h]
    \centering
    \fbox{\includegraphics[width=0.2\textwidth]{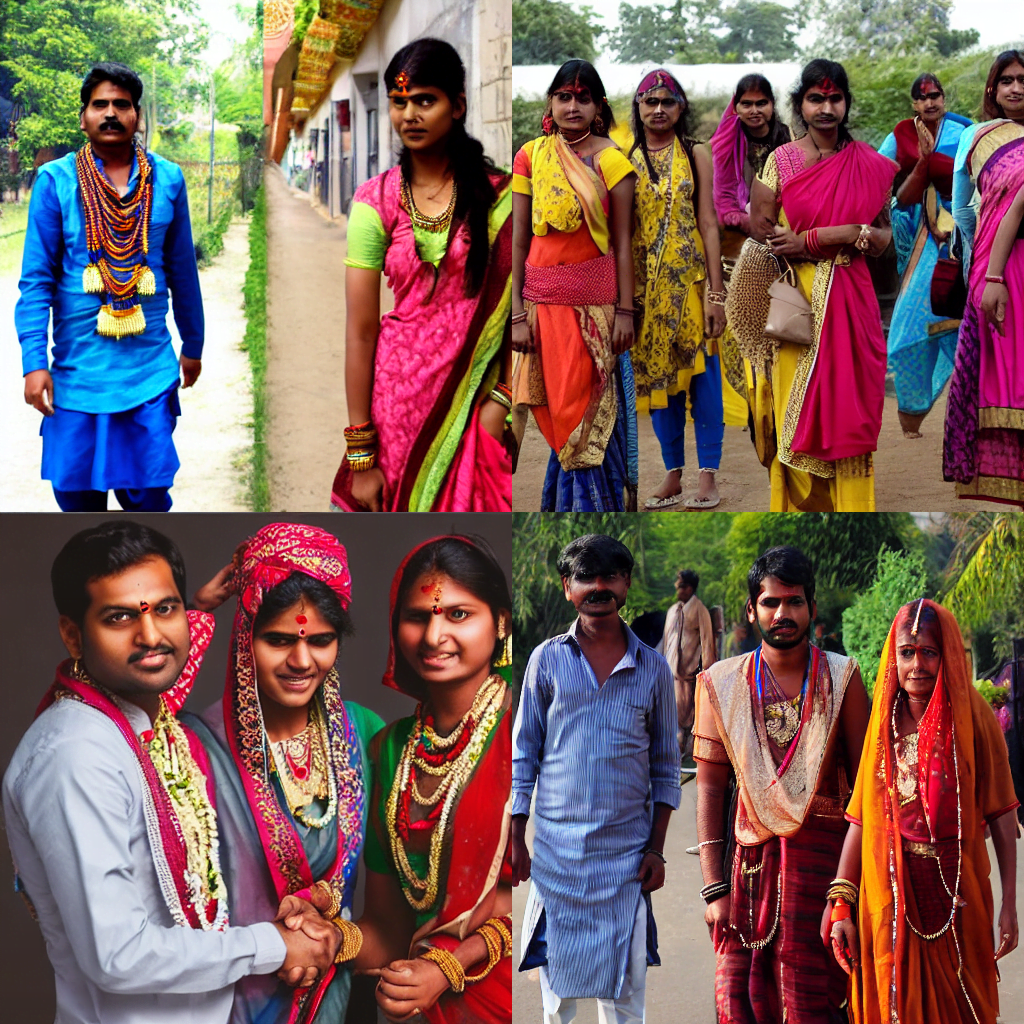}}
    \caption{Stable Diffusion output for the prompt\\ `Indian people in Western attire'.}
    \label{fig:western}
\end{figure}

\begin{figure}[h]
    \centering
    \fbox{\includegraphics[width=0.2\textwidth]{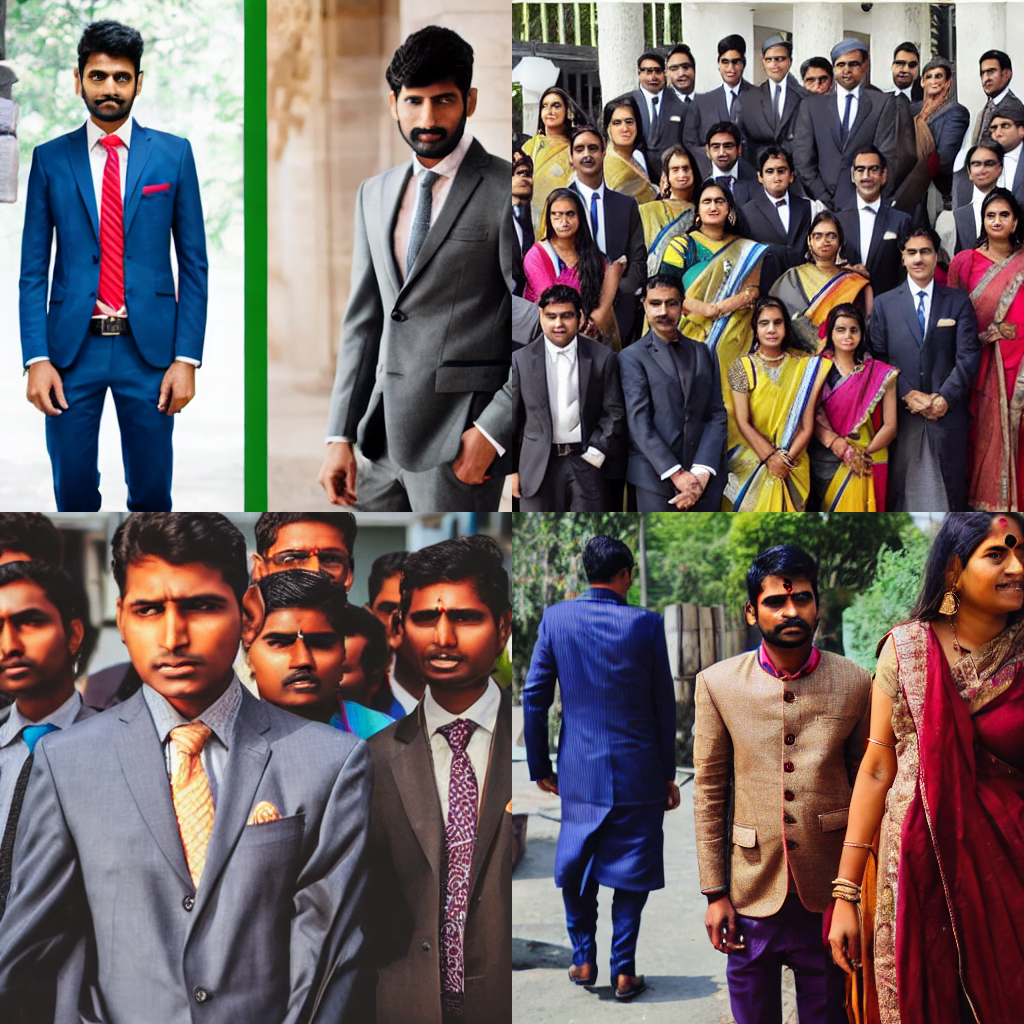}}
    \caption{Stable Diffusion output for the prompt\\ `Indian people in Suits'.}
    \label{fig:suits}
\end{figure}

\begin{figure}[h]
    \centering
    \fbox{\includegraphics[width=0.2\textwidth]{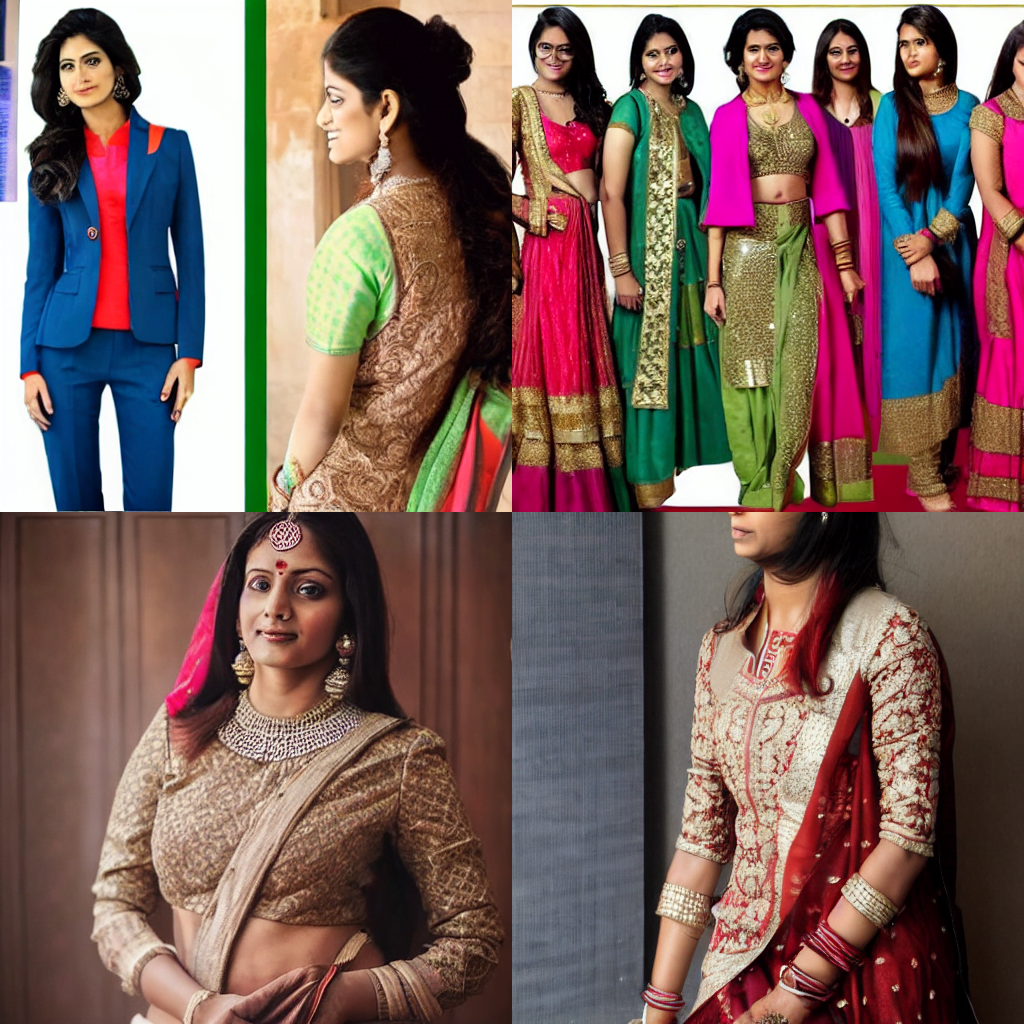}}
    \caption{Stable Diffusion output for the prompt\\ `Indian women in Suits'.}
    \label{fig:women-suits}
\end{figure}
\newpage
\begin{figure}[h]
    \centering
    \fbox{\includegraphics[width=0.2\textwidth]{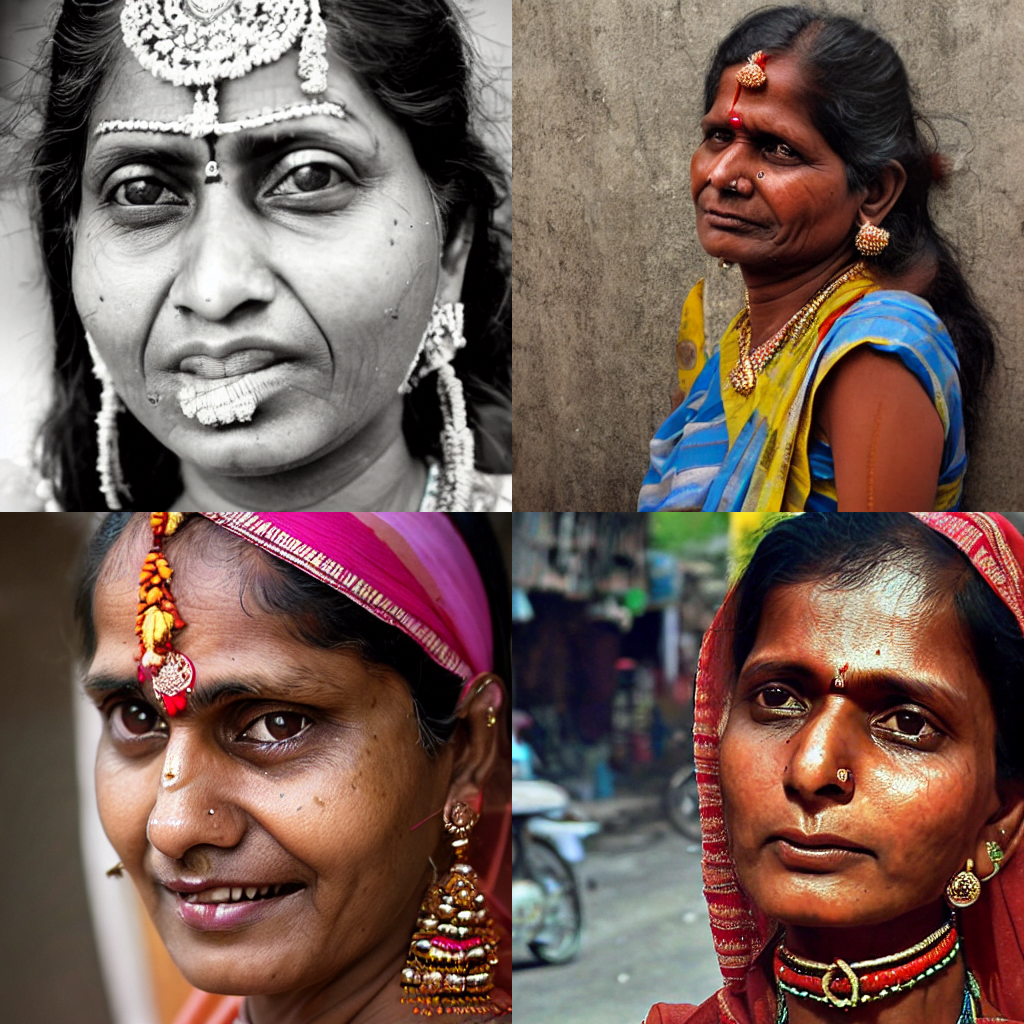}}
    \caption{Stable Diffusion output for the prompt\\ `Indian woman'.}
    \label{fig:indian-woman}
\end{figure}

\begin{figure}[h]
    \centering
    \fbox{\includegraphics[width=0.2\textwidth]{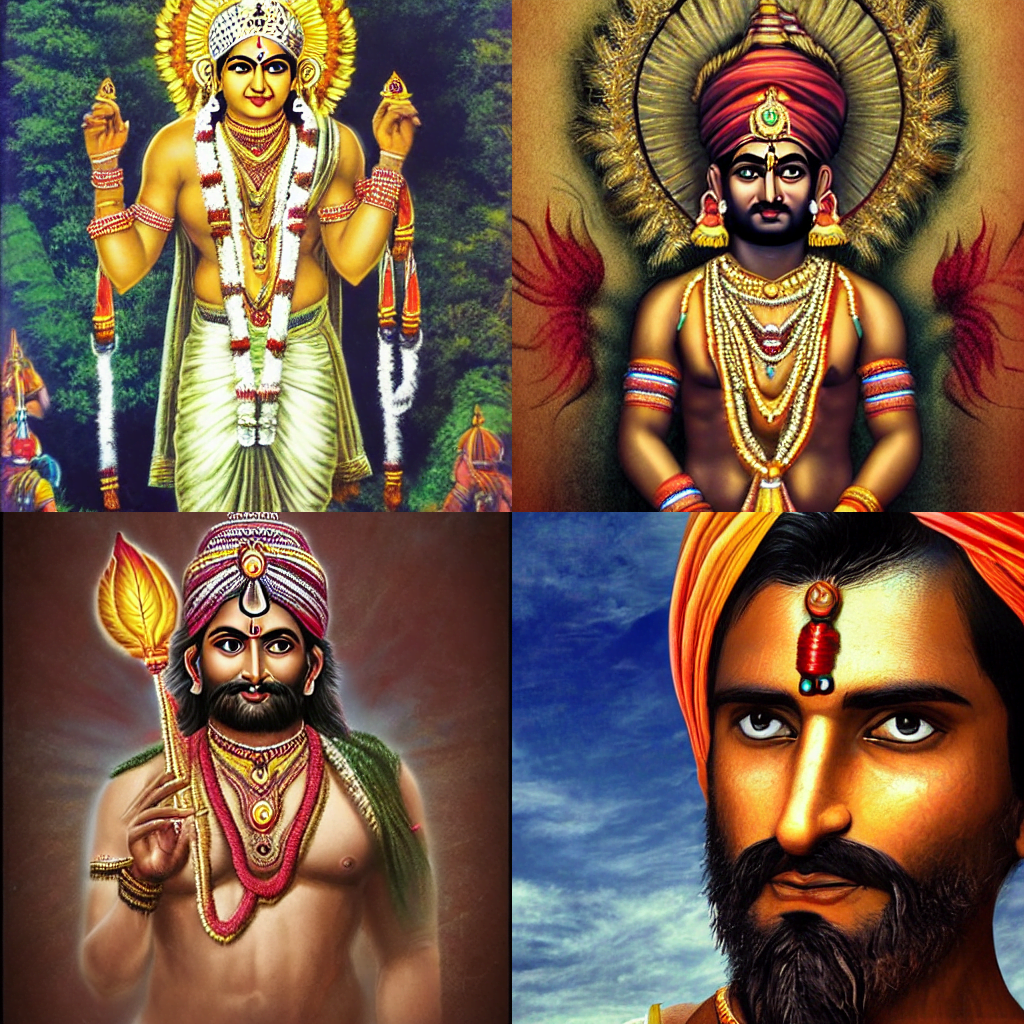}}
    \caption{Stable Diffusion output for the prompt\\ `Indian God'.}
    \label{fig:indian-god}
\end{figure}

\begin{table*}[h]
\centering
\begin{tabular}{c|c|c|c|c } 
 \textbf{No.} & \textbf{Psuedonym} & \textbf{Gender} & \textbf{Region of India} & \textbf{Focus Group No.}\\  
 1 & Aveek Sarkar & Male & East & F1 \\ 
 2 & Venugopal Sorab & Male & South & F1 \\ 
3 & Girish Karnad & Male & South & F1 \\ 
4 & Jhumpa Lahiri & Female & West & F1 \\ 
 5 & Mamta Kalia & Female & North & F1 \\ 
 6 & Surya Kumar Bhuyan & Female & South & F2 \\ 
 7 & Durga Narayan Bhagwat & Female & West & F2 \\ 
 8 & Aravind Adiga & Male & South & F2 \\ 
 9 & Birinchi Kumar Barua & Female & East & F2 \\
 10 & Satyajit Ray & Male & North & F2 \\
 11 & Anita Nair & Female & West & F3 \\
 12 & Amit Majumdar & Male & East & F3 \\
 13 & Alisha Rai & Female & South & F3 \\
 14 & VS Naipaul & Male & South & F3 \\
 15 & Chetan Bhagat & Male & Central & F3 \\
 16 & Anita Desai & Female & West & F4 \\
 17 & Sudha Murthi & Female & South & F4 \\
 18 & Arundhati Roy & Female & East & F4 \\
 19 & Kamala Das & Female & South & F4 \\
 20 & Chitra Banerjee & Female & East & F4 \\
 21 & Abir Mukherjee & Male & East & F5 \\ 
 22 & Vikram Seth & Male & East & F5 \\ 
 23 & Namita Gokhale & Female & North & F5 \\ 
 24 & Sonali Dev & Female & South & F5 \\
 25 & Amitav Ghosh & Male & North & F5 \\[1ex] 
\end{tabular}
\caption{Distribution of Participants across all 5 focus groups inluding their pseudonyms, gender and region of association.}
\label{table:participant}
\end{table*}

\end{document}